\documentclass[aoas,preprint]{imsart}

\RequirePackage[OT1]{fontenc}
\RequirePackage[numbers]{natbib}

\usepackage{amsthm,amsmath}
\usepackage{amsbsy}

\input xy
\xyoption{all}

\usepackage{algorithm}
\usepackage[noend]{algpseudocode}

\usepackage{epsf}
\usepackage{epsfig}

\pdfoutput=1

\startlocaldefs

\newcommand{\ci}{\perp\!\!\!\perp}
\newcommand{\nci}{\not\!\perp\!\!\!\perp}

\endlocaldefs

\begin{document}

\begin{frontmatter}

\title{On the analysis of personalized medication response and classification of case vs control patients in mobile health studies: the mPower case study}



\author{\fnms{} \snm{Elias Chaibub Neto, Thanneer M Perumal, Abhishek Pratap, Brian M Bot, Lara Mangravite, Larsson Omberg}\ead[label=e1]{elias.chaibub.neto@sagebase.org, Sage Bionetworks}}


\runauthor{}

\begin{abstract}
In this work we provide a couple of contributions to the analysis of longitudinal data collected by smartphones in mobile health applications. First, we propose a novel statistical approach to disentangle personalized treatment and ``time-of-the-day" effects in observational studies. Under the assumption of no unmeasured confounders, we show how to use conditional independence relations in the data in order to determine if a difference in performance between activity tasks performed before and after the participant has taken medication, are potentially due to an effect of the medication or to a ``time-of-the-day" effect (or still to both). Second, we show that smartphone data collected from a given study participant can represent a ``digital fingerprint" of the participant, and that classifiers of case/control labels, constructed using longitudinal data, can show artificially improved performance when data from each participant is included in both training and test sets. We illustrate our contributions using data collected during the first 6 months of the mPower study.
\end{abstract}






\end{frontmatter}

\section{Introduction}

Smartphones offer a unique opportunity to develop large scale studies of human health\cite{mcconnell2017, chan2017}. Features extracted from data collected by accelerometers, microphones, and touch screen sensors can provide objective measurements of human health and disease. Mobile health technologies may provide an unprecedented opportunity to improve overall management of global disease burden and health care costs. First, they may support the application of specialized medical care to remote populations that were previously under-served due to lack of local availability. Second, they may enable personalization of treatments in a manner that can safely and effectively meet individual medical needs while minimizing unintended negative outcomes, thereby reducing the overall burden of health care. In particular, smartphones can be used to determine if a patient is likely responding to its medication\cite{chaibubneto2016,chaibubneto2017}.

In the mPower (mobile Parkinson's observatory for worldwide evidence-based research) study, a participant is asked to perform distinct activity tasks, including speeded tapping, voice, gait, and balance tests\cite{bot2016}. Raw sensor data collected from each task is processed into activity specific features. Because the activity tasks are performed by the patient on a daily basis, before and after medication, over a long period of time, the processed data corresponds to time series of feature measurements, annotated according to whether the measurement was taken before or after the participant has taken medication. For each activity task, we extract several features.

Here, we describe a statistical approach to disentangle putative treatment effects from putative ``time-of-the-day" effects in observational studies, and apply it to data collected in the first 6 months of the mPower study. In this application, we are mostly interested in detecting the effect of levadopa on the participant's performance in an activity task (as measured by the difference in the participant's performance when medicated in comparison to when the participant is unmedicated). But, because it is the participant who decides when he/she will perform the activity task, it is possible that any observed difference in performance might be due to circadian rhythms or daily routine activities (as well as, to long term confounders) instead of due to the medication.

We use the recorded time-of-the-day that the activity was performed as a surrogate variable for the short term cyclic potential influences and model treatment and time-of-the-day effects under the assumption of no unmeasured confounders. (In a previous contribution we propose the use of instrumental variables in randomized experiments with imperfect compliance as an approach to account for unmeasured confounders\cite{chaibubneto2017}.) The present approach is based on conditional independence tests implemented via t-tests in linear regression models, and represents an improvement over another previous contribution\cite{chaibubneto2016}. We discuss how residual autocorrelation can inflate (or deflate) the t-test p-values, and we propose the use of ARIMA processes and heterocedasticity and autocorrelation consistent estimators of covariance as remedies to these issues. Finally, we describe how to combine the results of the feature specific conditional independence tests into a single union-intersection testing procedure.

Our second contribution, is to describe an interesting phenomenon observed in the mPower data, namely, that the sensor data collected from a study participant provides a digital fingerprint of the participant. This observation has important practical consequences for any analyses performed at a population level. In particular, for the classification of Parkinson's diasease (PD) cases versus healthy controls one needs to be careful when designing the training/test random split used for building the case/control classifier, since if data from each participant is included in both the training and test set, then the classifier performance can be artificially improved because the digital fingerprint makes it easy to discriminate one participant from another. We start by illustrating this issue in detail.

\section{On the classification of PD patients and healthy controls using smartphone sensor data: the digital fingerprint phenomenon}

Several smartphone based studies have focused in Parkinson's disease\cite{arora2014, arora2015, stephen2015, trister2016, zhan2016, chaibubneto2016}. Current tests for the diagnosis of PD are based on subjective neurological examinations performed in-clinic (although the use of objective measurements collected by smartphones has been proposed as a potential alternative approach to the remote characterization of PD\cite{arora2014, arora2015}).

\begin{figure}[!h]
\begin{center}
\includegraphics[width=\linewidth]{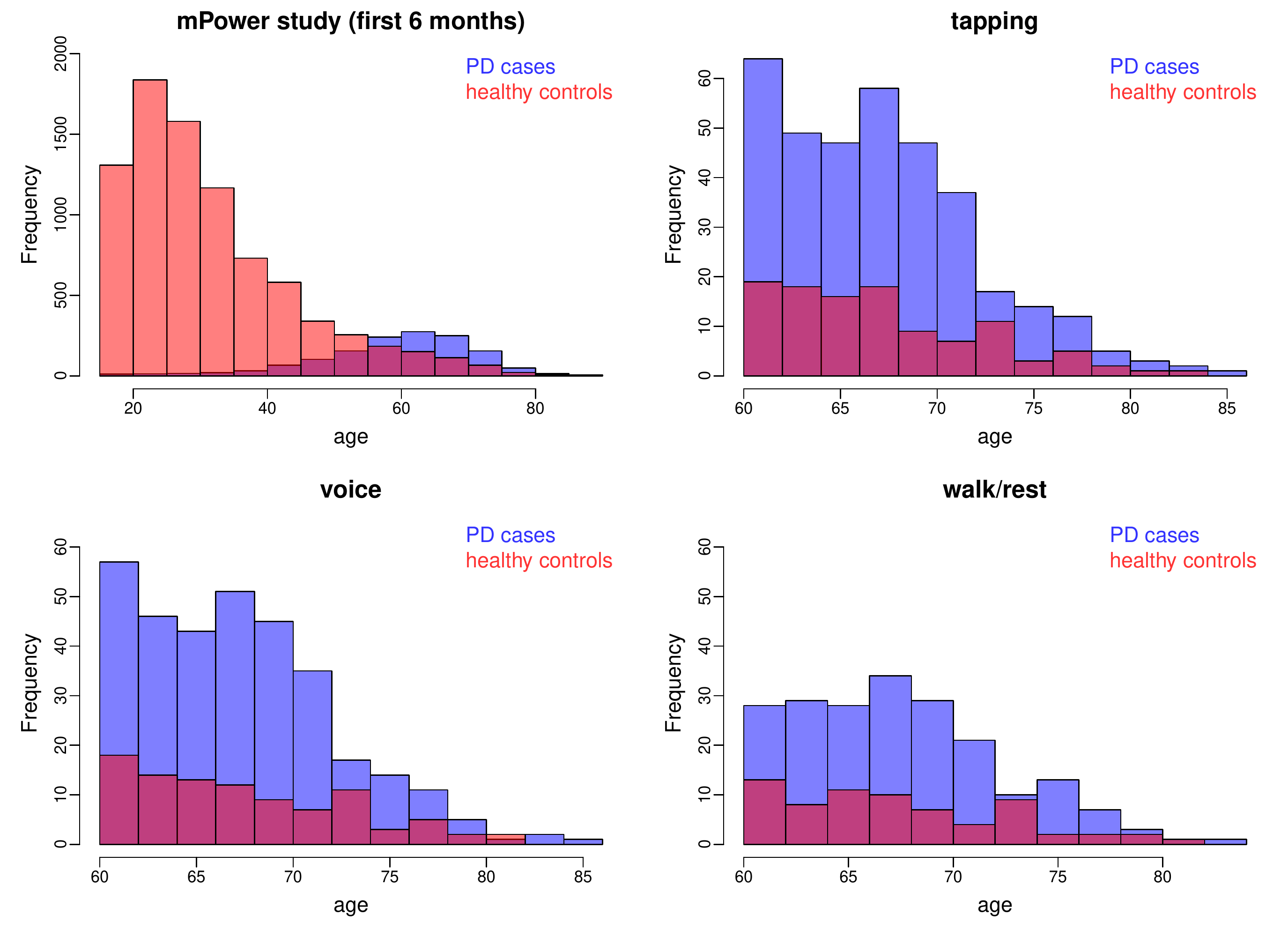}
\caption{The top left panel shows age distributions for PD cases (blue) and healthy controls (red) in the first 6 months of the mPower study. Due to the poor overlap of the distributions (note the much younger control population), we focus our analyses on a subset of participants with age distributions spread over similar ranges (namely, between 61 and 86 years). For the tapping activity our analyses where restricted to 315 cases and 109 controls, while the voice analyses focused on 277 cases and 96 controls, and the walk (and rest) analyses employed 170 cases and 69 controls. (The distributions are different for the different tasks because participants perform the distinct tasks with different frequencies.)}
\label{fig:agematched}
\end{center}
\end{figure}

Here, we investigate the accuracy of the classification of PD participants versus healthy controls in data collected during the first 6 months of the mPower study. Because each participant contributes several samples to the study, we end up with longitudinal data for each extracted feature. We analyzed each activity task separately, using a subset of PD cases and healthy controls with age distributions spread over similar ranges (see Figure \ref{fig:agematched}).

\begin{figure}[!h]
\begin{center}
\includegraphics[width=\linewidth]{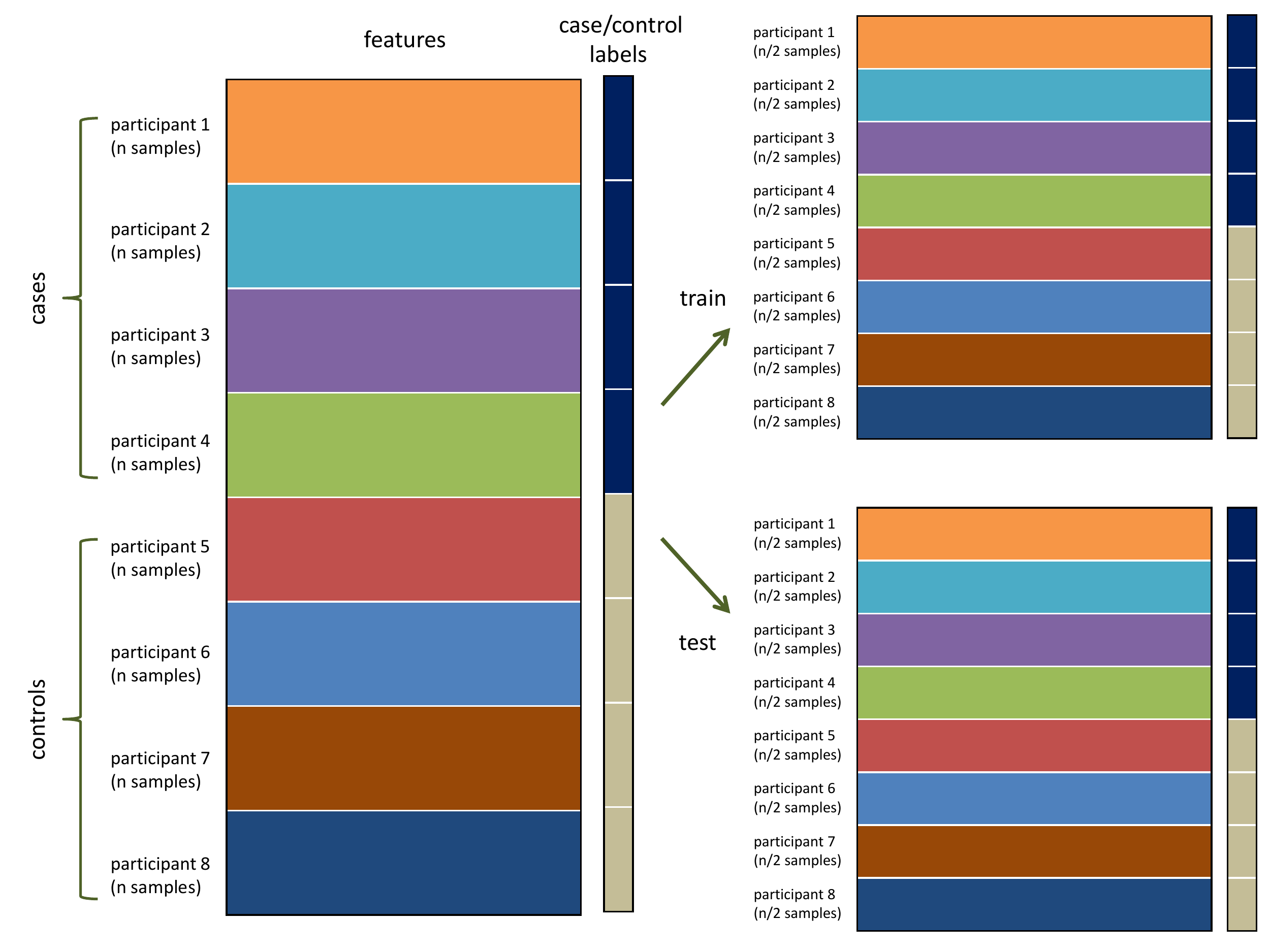}
\caption{Train/test split where the data contributed by each participant is included in both the training and test set. The schematic represents a data set composed of 4 PD cases (participants 1 to 4), and 4 healthy controls (participants 5 to 8). Each participant has longitudinal data from n activity tasks, and the full feature data matrix is composed of 8n rows/samples and p columns/features. The vector of case/control labels has 8n elements, with PD labels shown in dark and healthy control labels shown in light color. The full data set is randomly split in two equally sized training and test sets, with both sets containing approximately n/2 samples from each one of the participants.}
\label{fig:wrongsplit}
\end{center}
\end{figure}

\begin{figure}[!h]
\begin{center}
\includegraphics[width=\linewidth]{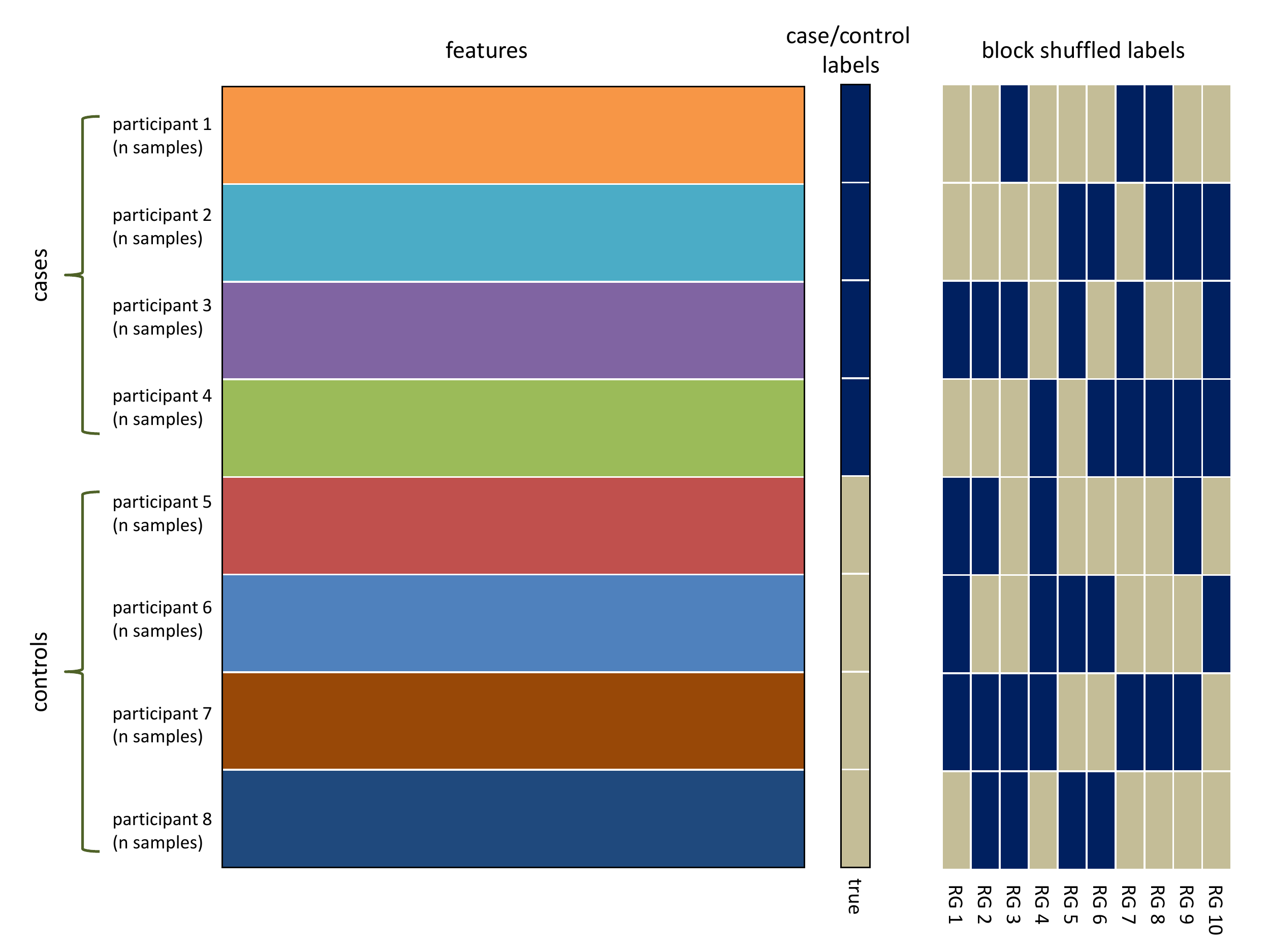}
\caption{Generation of random groups of participants by blocked shuffling of the case and control labels. The schematic shows 10 distinct examples (RG 1 to RG 10) of arbitrarily chosen groups of participants generated by randomly shuffling the participant's labels as a block. For instance, the first random group (RG 1) was generated by randomly shuffling the original labels (as blocks) so that the labels of participants 1, 2, and 4 changed from ``case" to ``control", the labels of participants 5, 6, and 7 changed from ``control" to ``case", and the labels of participants 3 and 8 remained the same. (Note that for each participant the labels are changed across all n samples.) For each random group, the data is split into training and test sets, as described in Figure \ref{fig:wrongsplit}, and a classifier is trained to predict the random group labels.}
\label{fig:blockshuffle}
\end{center}
\end{figure}

\begin{figure}[!h]
\begin{center}
\includegraphics[width=2.4in]{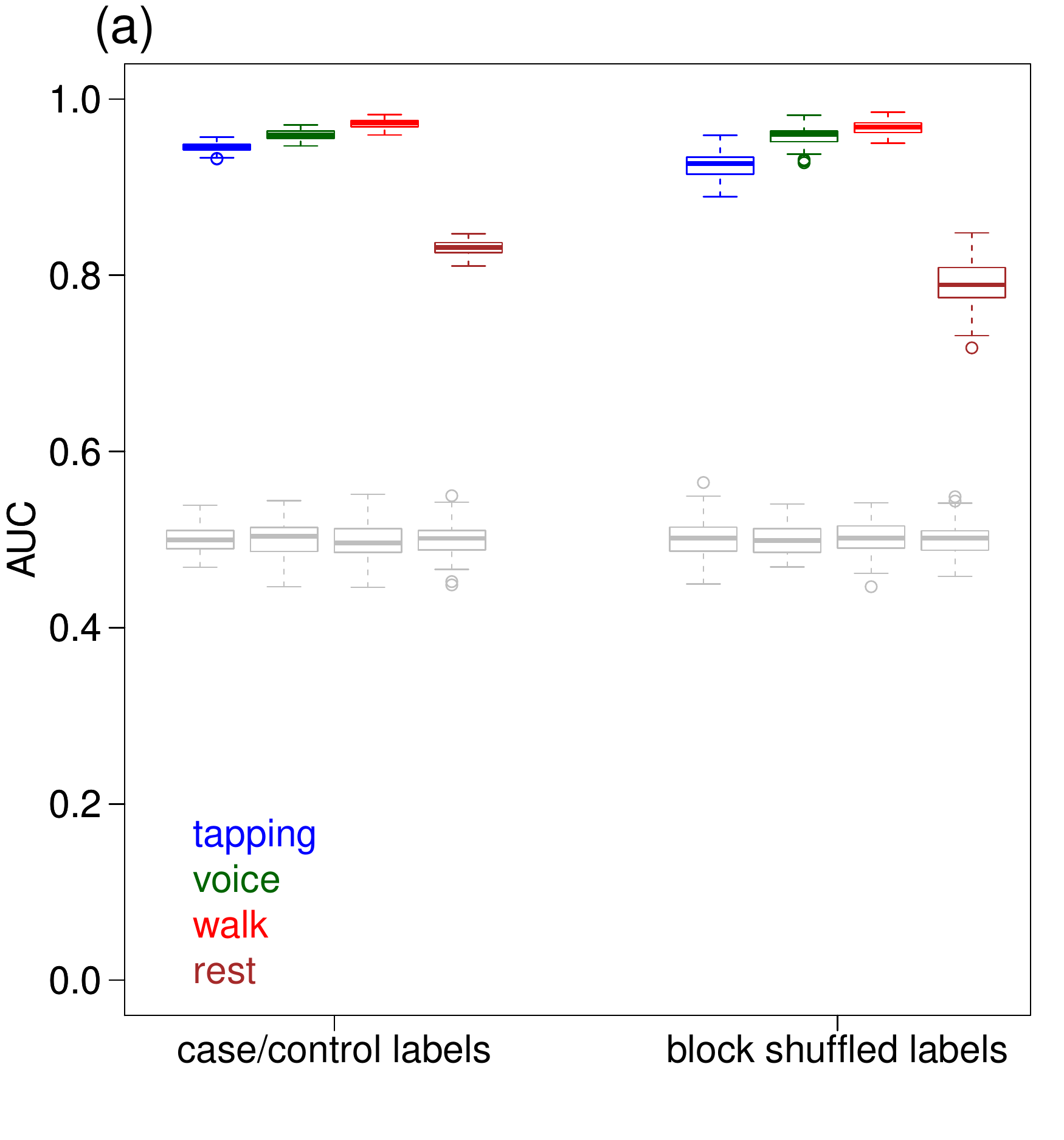}
\includegraphics[width=2.4in]{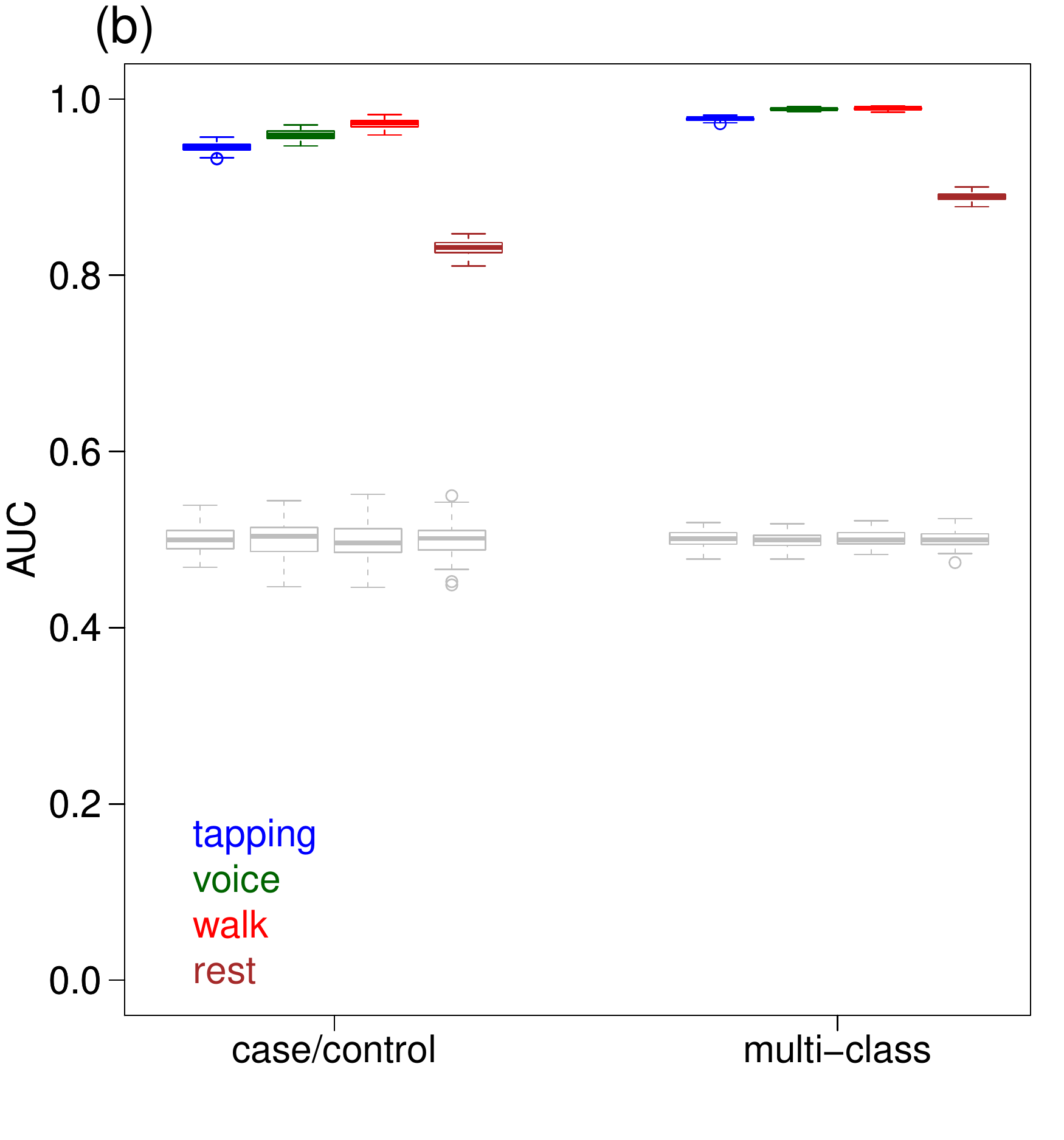}
\caption{Panel a shows the performance comparison of the case/control (original labels) versus random groups (block shuffled labels) classification, when data contributed by each participant is included in both the training and test set. We analyzed the tapping, voice, walk, and rest activities, separately (employing 41, 116, 19, and 13, tap, walk, rest, and voice features, respectively). The PD cases and healthy controls were randomly selected from the larger set of approximately age matched participants described in Figure \ref{fig:agematched}. Panel b shows the comparison of the classification AUC in the multi-class problem (classifying the participant identities) versus the binary classification of cases/controls. Results based on the train/test random split described in Figure \ref{fig:wrongsplit}. For the multi-class AUC computation, we adopted the one-vs-one approach\cite{handTill2001}. In both panels, the grey boxplots represent the results for completely shuffled labels across all samples (where the same participant can have samples labeled as cases and controls).}
\label{fig:blockshuffleresults2}
\end{center}
\end{figure}

\begin{figure}[!h]
\begin{center}
\includegraphics[width=\linewidth]{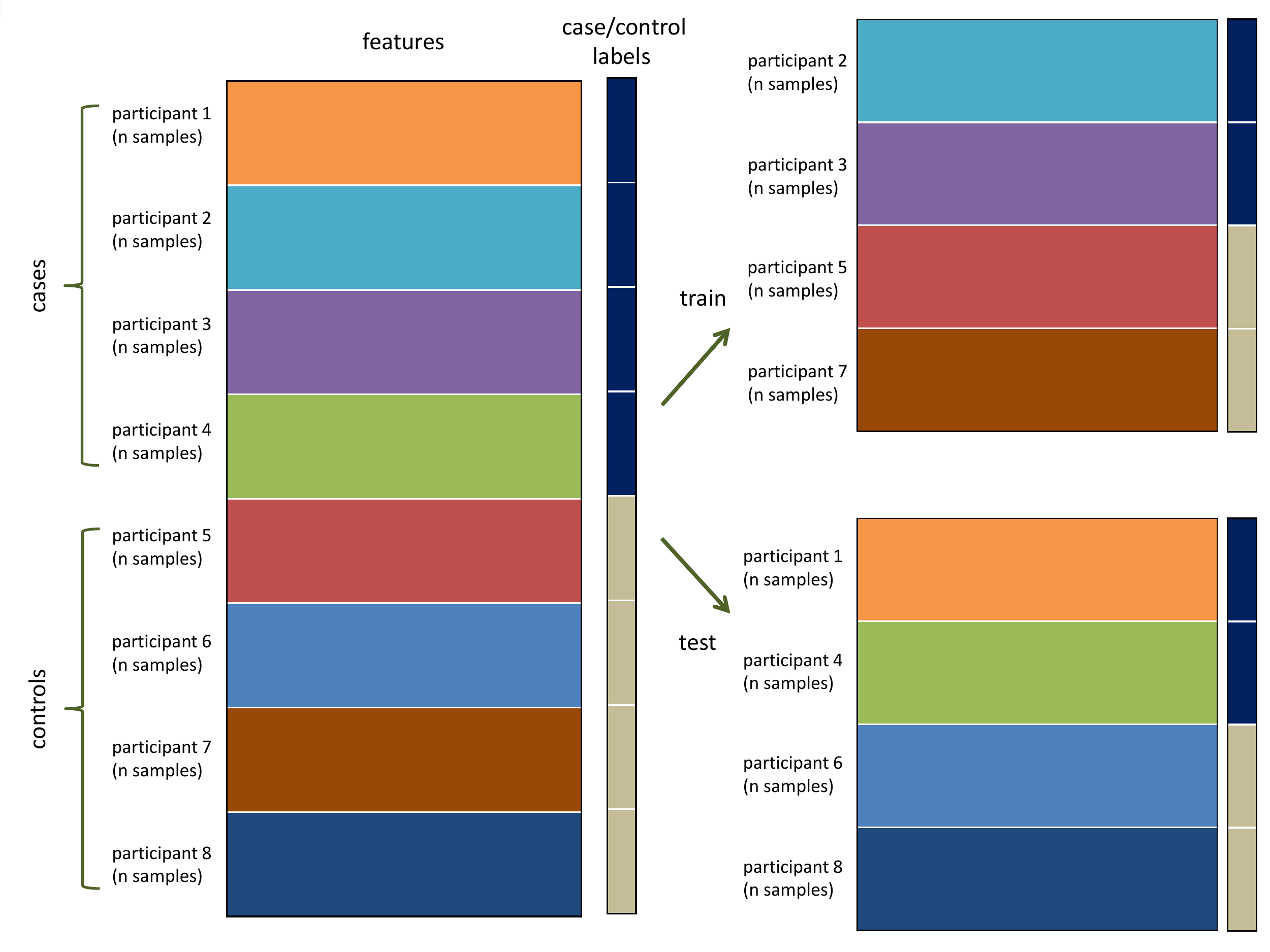}
\caption{Train/test split where the data contributed by each participant is either included in the training or in the test set. The schematic represents a data set composed of 4 PD cases (participants 1 to 4), and 4 healthy controls (participants 5 to 8). Each participant has data from n activity tasks, and the full feature data matrix is composed of 8n rows/samples and p columns/features. The vector of case/control labels has 8n elements, with PD labels shown in dark and healthy control labels shown in light color. The data of each participant is randomly assigned to either the training or test set. In this example, the training set received all the data (n samples) of participants 2, 3, 5, and 7, while the test set receives the data of participants 1, 4, 6, and 8.}
\label{fig:correctsplit}
\end{center}
\end{figure}

\begin{figure}[!h]
\begin{center}
\includegraphics[width=1.6in]{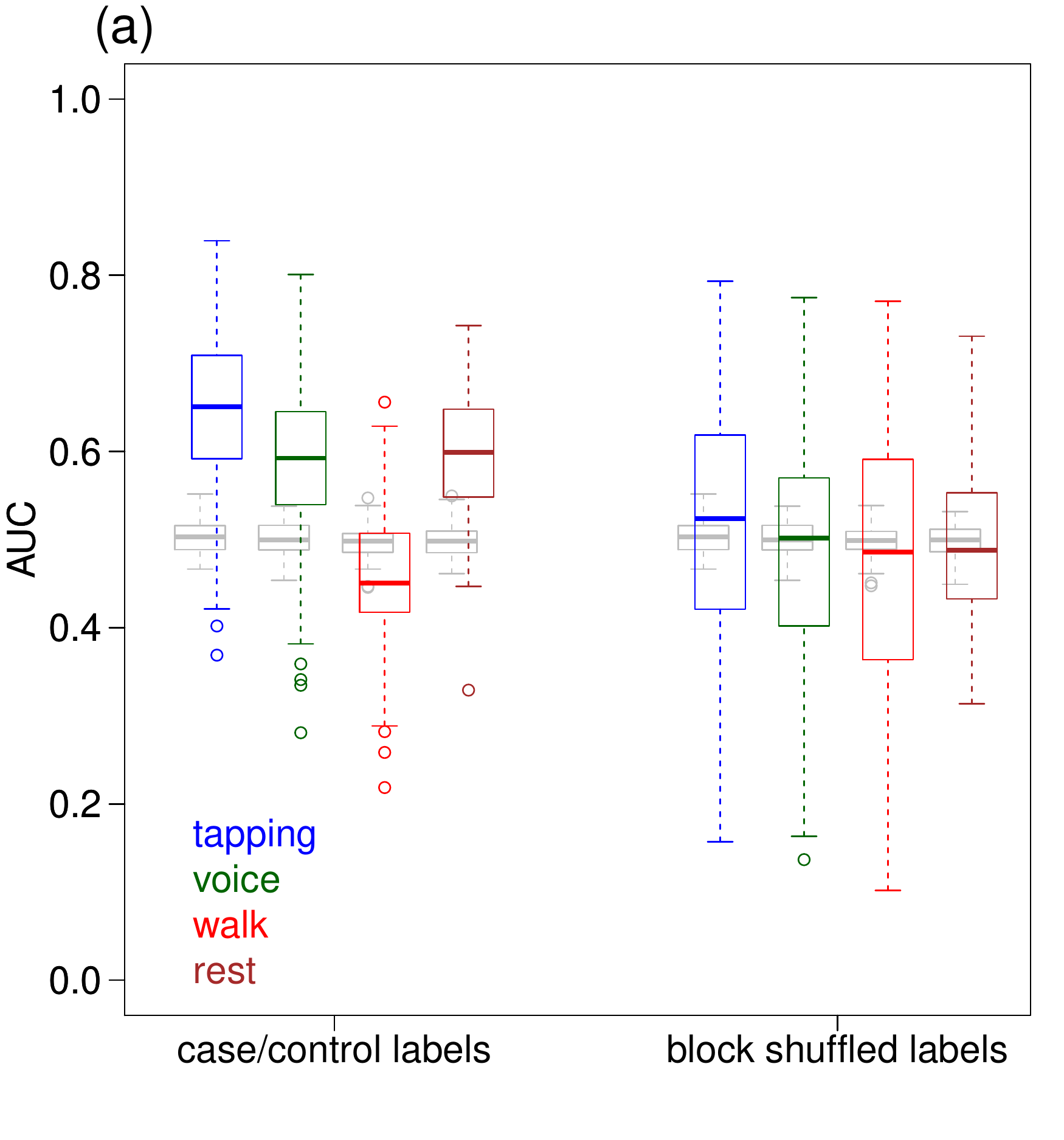}
\includegraphics[width=1.6in]{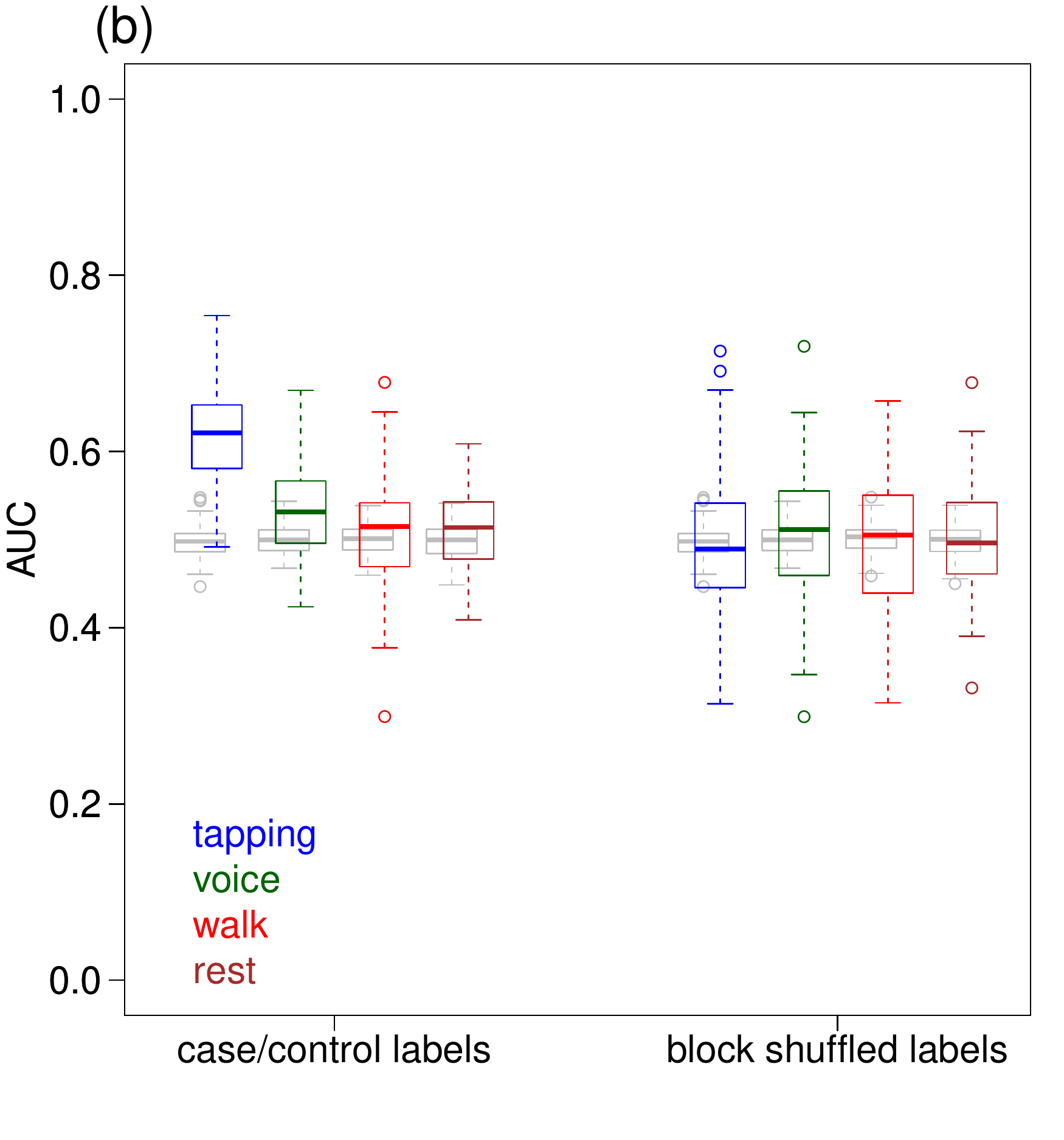}
\includegraphics[width=1.6in]{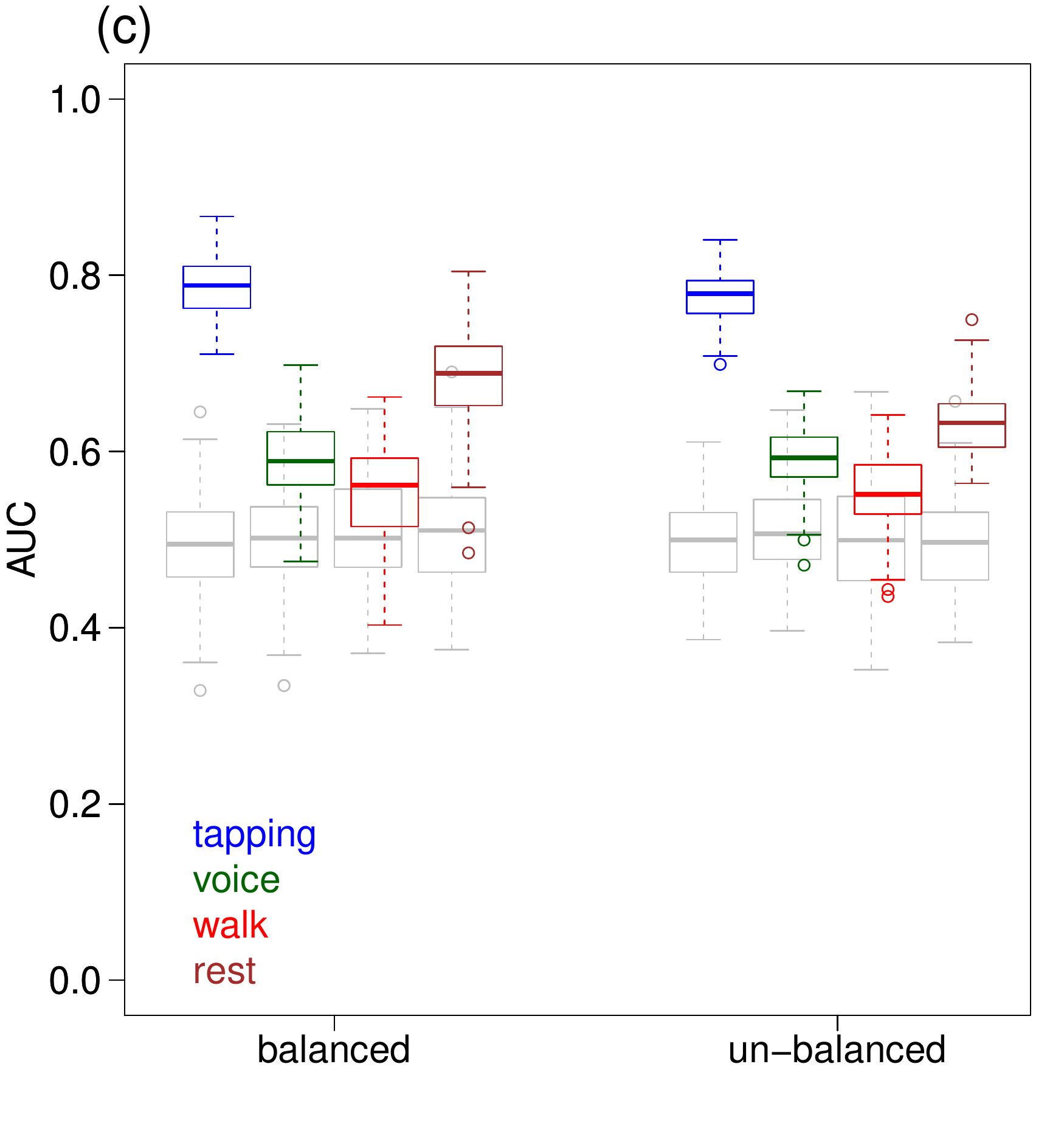}
\caption{Panel a shows the performance comparison of the case/control (original labels) versus random groups (block shuffled labels) classification, when the data contributed by each participant is either included in the training or in the test set. Results are based in the same data analyzed in Figure \ref{fig:blockshuffleresults2}a (i.e., 10 cases, 10 controls with 100 samples from each), but using the train/test split described in Figure \ref{fig:correctsplit}. Panel b shows the same comparison, but based in a larger number of participants (23 PD cases and 23 healthy controls), but a smaller amount of longitudinal data (44 samples per participant). Note the smaller spread of the boxplots in comparison to the results in panel a. Panel c shows the classification performance based on larger number of participants, but using a single ``collapsed sample" per participant. The figure reports results for balanced and un-balanced cases. In all panels, the grey boxplots represent the results for completely shuffled labels across all samples.}
\label{fig:correctsplitresults2}
\end{center}
\end{figure}

We found out that smartphone data collected from a given study participant can represent a ``digital signature", or a ``digital fingerprint", of the participant. As a consequence, one needs to be careful when designing the train/test split of the data used to train a case/control classifier. Explicitly, if data from each participant is included in both the training and test set, as illustrated in Figure \ref{fig:wrongsplit}, then the classifier performance can be artificially improved because it is easy to discriminate one participant from another. In other words, the classifier is able to learn about individual characteristics of the participants to a greater extent than about the differences between cases and controls. To illustrate this point we apply a random forest classifier\cite{breimam2001} to a randomly chosen (and balanced) subset of the approximately age matched data from Figure \ref{fig:agematched} in order to classify not only cases versus controls, but also distinct random groups of participants, generated by block shuffling the case/control labels as described in Figure \ref{fig:blockshuffle}.

Figure \ref{fig:blockshuffleresults2}a presents the performance of the random forest classifier according to the area under the receiver operating characteristic curve (AUC). For each one of the separate activity tasks, the data was composed of 10 randomly selected PD cases, 10 randomly selected healthy controls, and features extracted from 100 randomly selected activity tasks per participant. The boxplots in Figure \ref{fig:blockshuffleresults2}a represent the results of 100 random splits of the data into equally sized training and test sets, as described in Figure \ref{fig:wrongsplit}. Note that for the boxplots in the right (``block shuffled labels"), each one of the train/test random splits used a distinct version of the block shuffled labels, whereas the boxplots in the left (``original labels") used the same case/control original labels in all 100 train/test random splits. (We used the same 100 random train/test splits in the analysis of the original and block shuffled labels.) The figure clearly shows that the performance of the case/control classification is comparable to the random group classifications, showing that a random forest classifier is equally able to discriminate between any two arbitrarily chosen groups of participants, as between PD cases and healthy controls. This observation suggests that the classifiers are mostly learning the individual characteristics of each participant, or, in other words, the digital fingerprints of the participants. To further corroborate this point we also employed a random forest to evaluate our ability to classify the participants in a multi-class classification problem, where the classes correspond to the participant identities. Figure \ref{fig:blockshuffleresults2}b shows that the AUCs of the multi-class classifier were slightly higher than of the binary classification, even though the multi-class classification is a much harder problem (recall that because we have 20 participants, the probability of correctly guessing the identity of a participant is only 0.05, while in the binary classification problem we can guess the correct class half of the time on average).

Note that while the training/test split described in Figure \ref{fig:wrongsplit} is perfectly valid from the machine learning perspective (as the classifier is trained and evaluated in completely separated data sets), the digital fingerprint phenomenon creates, nonetheless, a serious artifact, and clearly illustrates that one should avoid including data from each participant in both training and test sets. Observe, as well, that the digital fingerprint of a participant may arise not only from biological characteristics of a participant, but also from non-biological artifacts. For instance, it is likely that non-biological artifacts such as the length of the room where a participant performs the walking task might influence the extracted features in the walking activity (since the shorter the room, the larger the number of turns a participant will need to do during the 30 seconds of duration of the task, and it is not unreasonable to expect that the acceleration patterns will pick up these differences).

In order to circumvent the digital fingerprint issue, we adopt a train/test split where the data contributed by each participant is either included in the training or in the test set, as described in Figure \ref{fig:correctsplit}. Figure \ref{fig:correctsplitresults2}a reports the performance of the case/control classification using the exact same data as in Figure \ref{fig:blockshuffleresults2}a, but adopting the train/test random split described in Figure \ref{fig:correctsplit}. It clearly shows weaker average classification performances and considerably larger variability, in comparison to the results in Figure \ref{fig:blockshuffleresults2}a. This observation is, nonetheless, not surprising given that the classification was based on a small number of participants contributing a large number of samples. As expected, results tend to be less variable as we increase the number of participants and decrease the number of samples per participant, as illustrated in Figure \ref{fig:correctsplitresults2}b, where results were based on 23 PD cases and 23 healthy controls, each one contributing 44 samples (as opposed to 10 cases and 10 controls contributing 100 samples each).

Figure \ref{fig:correctsplitresults2}c reports results using all the ``age matched" participants from Figure \ref{fig:agematched}, but a single ``collapsed sample" per participant (namely, for each feature of each participant, we used the median of the feature values across all longitudinal samples contributed by the participant as the collapsed feature value). Note that, in this case, the random train/test split described in Figure \ref{fig:correctsplit} reduces to the standard train/test split for data sets containing a single sample per participant. The figure reports results for both balanced and un-balanced cases. For the tapping, voice, and walk/rest activities, the un-balanced analysis was based, respectively, on 315 cases/109 controls, 277 cases/96 controls, and 170 cases/69 controls. For the balanced analyzes, we randomly selected the number of PD case participants to match the respective number of healthy controls. The results show that while there is signal in the data to discriminate between cases and controls, the accuracy of the classification is not particularly high.

Overall, our investigations suggest that, at least for clinical studies such as the mPower that are run ``in the wild", a much larger number of participants and/or more sophisticated machine learning algorithms are necessary in order to achieve state-of-the-art performance in the discrimination of PD cases versus healthy controls using smartphone sensor data.

In the next section, we switch to our methodological contribution.

\section{On the analysis of personalized medication response}

\begin{figure}[!h]
\centering
\includegraphics[width=\linewidth]{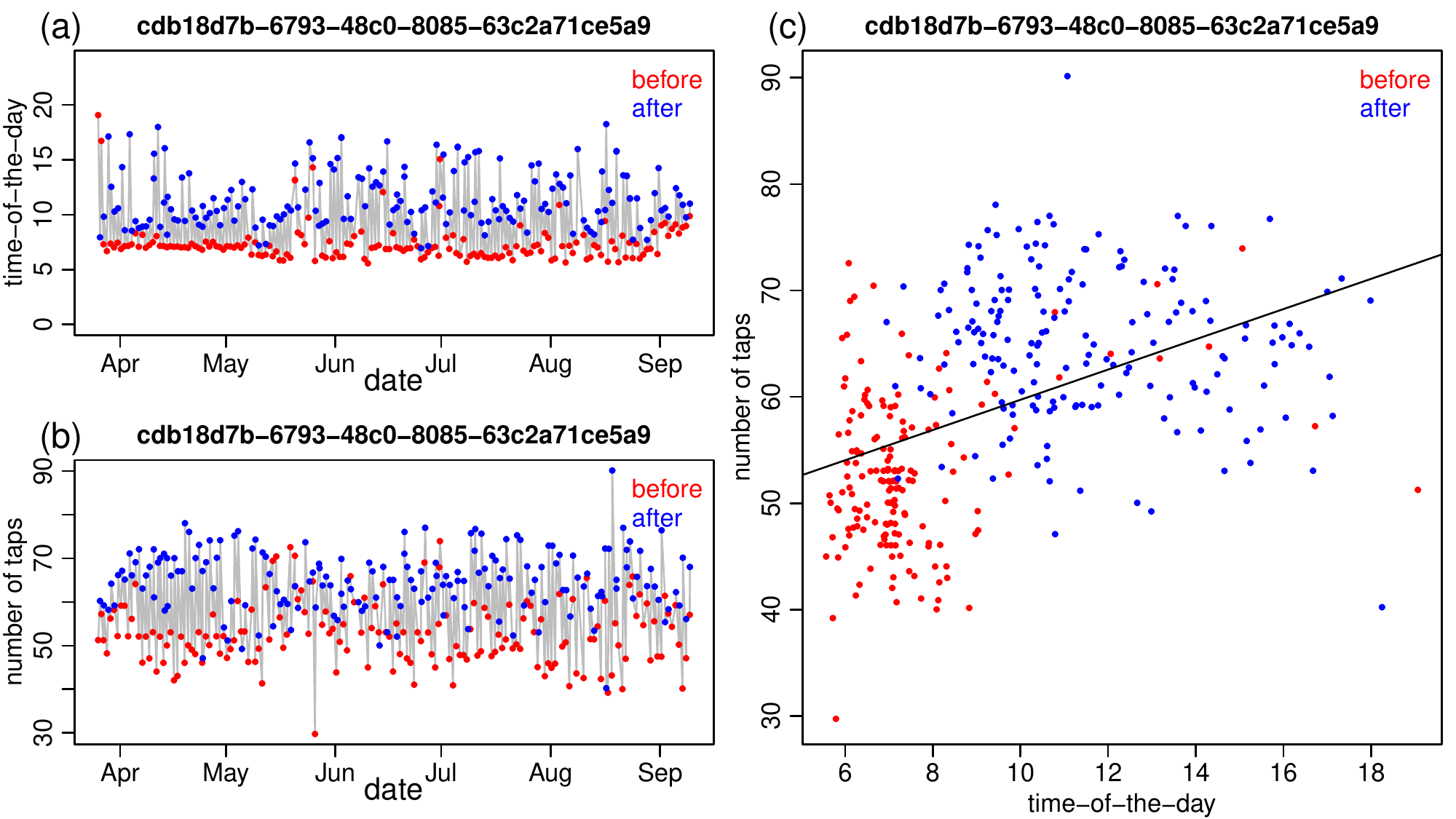}
\caption{\textbf{Marginal associations between treatment (before/after medication status), time-of-the-day and number of taps, for one study participant.} Panel a shows that the participant usually performs the before medication tapping tasks (red dots) earlier in the day than the after medication tasks (blue dots). Panel b shows the participant also tends to achieve better performance (larger number of taps) in tasks performed after medication. Panel c, nonetheless, also shows that large number of taps tends to be associated with later times. Hence, it is possible that the medication and/or circadian rhythms/daily routine activities might be responsible for the difference in performance between the before and after medication tapping tasks observed in this participant.}
\label{fig:confoundingbytod}
\end{figure}

We are interested in determining whether a particular patient is responding to medication, that is, we want to determine if the treatment $X$ has an effect on a outcome $Y$. Our focus is on the personalized level, as opposed to the traditional focus of medicine, where the goal is to establish treatment efficacy at a population level for a target cohort of patients\cite{topol2012,schork2015}. However, since mPower is an observational study, the associations observed between treatment and outcome measurements might be due to unmeasured confounders, and it is not possible to conclude with certainty that a difference in performance between before/after medication tasks is actually due to the medication. In particular, causal inferences at the personalized level are especially vulnerable to confounding effects that arise in a cyclic fashion over the day (such as circadian rhythms and daily routine activities). For instance, we observed in our data (Figure \ref{fig:confoundingbytod}a) that some participants tended to perform the ``before medication" activity tasks earlier in the day than the ``after medication" tasks. For these participants, it is not possible to conclude that an observed improvement in performance between tasks performed before versus after medication are suggestive of a medication effect, since the difference in performance might be due to daily cyclic confounders (Figures \ref{fig:confoundingbytod}b and c).

Fortunately, the time-of-the-day that the activity is performed is always recorded by the smartphone, and we can use it as a surrogate variable for circadian rhythms and daily routine confounders in our analyses. Hence, the available data for the analyses corresponds to triplets, $\{X, T, Y\}$ (where $T$ stands for the time of the day that the activity task was performed). In the following we describe how we can use a subset of the conditional independence relationships spanned by the $\{X, T, Y\}$ variables in order the determine whether a difference in performance might be due to a putative treatment or putative ``time-of-the-day" effect (or both).

\subsection{Disentangling putative treatment effects from putative ``time-of-the-day" effects}

Under the assumption that $X$, $T$, and $Y$ are not influenced by unmeasured confounders (in addition to other standard causal graphical model assumptions mentioned at the end of this section), it is possible to use a subset of the conditional independence relationships spanned by the $\{X, T, Y\}$ variables to determine if $X$ has a causal effect on $Y$, or if $T$ has a causal effect on $Y$, or if both $X$ and $T$ have causal effects on $Y$, irrespective of the causal relationship between $X$ and $T$.

Explicitly, consider the putative causal models listed below.
\begin{equation}
{\scriptsize
\xymatrix@-0.5pc{
_{M_1} & *+[F-:<10pt>]{T} & & & _{M_2} & *+[F-:<10pt>]{T} \ar[dr] & & & _{M_3} & *+[F-:<10pt>]{T} \ar[dr] & \\
 *+[F-:<10pt>]{X} \ar[ur] \ar[rr] && *+[F-:<10pt>]{Y} && *+[F-:<10pt>]{X} \ar[ur] && *+[F-:<10pt>]{Y} && *+[F-:<10pt>]{X} \ar[ur] \ar[rr] && *+[F-:<10pt>]{Y} \\
_{M_4} & *+[F-:<10pt>]{T} \ar[ld] & & & _{M_5} & *+[F-:<10pt>]{T} \ar[dl] \ar[dr] & & & _{M_6} & *+[F-:<10pt>]{T} \ar[dl] \ar[dr] & \\
 *+[F-:<10pt>]{X} \ar[rr] && *+[F-:<10pt>]{Y} && *+[F-:<10pt>]{X} && *+[F-:<10pt>]{Y} && *+[F-:<10pt>]{X} \ar[rr] && *+[F-:<10pt>]{Y} \\
_{M_7} & *+[F-:<10pt>]{T} & & & _{M_8} & *+[F-:<10pt>]{T} \ar[dr] & & & _{M_9} & *+[F-:<10pt>]{T} \ar[dr] & \\
 *+[F-:<10pt>]{X} \ar[rr] && *+[F-:<10pt>]{Y} && *+[F-:<10pt>]{X} && *+[F-:<10pt>]{Y} && *+[F-:<10pt>]{X} \ar[rr] && *+[F-:<10pt>]{Y} \\
}}
\label{eq:models}
\end{equation}
Models $M_1$ and $M_4$ are indistinguishable in terms of conditional independence relationships. In the language of graphical models\cite{Lauritzen1996}, $M_1$ and $M_4$ are Markov equivalent\footnote{A simple graphical criterion (Verma and Pearl 1990) for determining if two directed and acyclic graphs (DAGs) are Marvov equivalent is to inspect if the DAGs have the same skeleton and the same set of v-structures (where the skeleton of a DAG is obtained by replacing the directed edges by undirect ones, and a v-structure is composed by two converging arrows whose tails are not connected by an arrow). For instance, models $M_1$ and $M_4$ have the same skeleton,  $\xymatrix@-1.0pc{T \ar@{-}[r] & X \ar@{-}[r] & Y}$, and the same set of v-structures (namely, no v-structures), and we say that models $M_1$ and $M_4$ belong to the same equivalence class.}.  Note that while $M_1$ and $M_4$ differ with respect to the causal relation between $X$ and $T$ (where $X \rightarrow T$ in model 1, and $X \leftarrow T$ in model 4), both models represent a causal effect of the treatment on the outcome (i.e., $X \rightarrow Y$). Similarly, $M_2$ and $M_5$ are Markov equivalent and depict an effect of the time-of-the-day on the outcome, but no treatment effect, while $M_3$ and $M_6$ are Markov equivalent and represent the case where both treatment and time-of-the-day effects influence the outcome. Models $M_7$, $M_8$, and $M_9$ represent, respectively, the causal DAGs for treatment, time-of-the-day, and both effects in the case where $X$ and $T$ are not even associated.

The subset of the conditional independence relations that can be used to distinguish between the 6 equivalence classes of models, $\{M_1, M_4\}$, $\{M_2, M_5\}$, $\{M_3, M_6\}$, $M_7$, $M_8$, and $M_9$, is given by,
\begin{equation}
{\scriptsize
\begin{array}{ccccccl}
\hline
\mbox{Models}  & T,X & Y,X & Y,T & Y,X \mid T & Y,T \mid X & \mbox{Putative effect} \\
\hline
\{M_1, M_4\} & T \nci X & Y \nci X & Y \nci T & Y \nci X \mid T & Y \ci T \mid X & \mbox{treatment} \\
\{M_2, M_5\} & T \nci X & Y \nci X & Y \nci T & Y \ci X \mid T & Y \nci T \mid X & \mbox{time-of-the-day} \\
\{M_3, M_6\} & T \nci X & Y \nci X & Y \nci T & Y \nci X \mid T & Y \nci T \mid X & \mbox{both} \\
M_7 & T \ci X & Y \nci X & Y \ci T & Y \nci X \mid T & Y \ci T \mid X & \mbox{treatment} \\
M_8 & T \ci X & Y \ci X & Y \nci T & Y \ci X \mid T & Y \nci T \mid X & \mbox{time-of-the-day} \\
M_9 & T \ci X & Y \nci X & Y \nci T & Y \nci X \mid T & Y \nci T \mid X & \mbox{both} \\
\hline
\end{array}}
\label{eq:ciarray}
\end{equation}
where we adopt the notation $\ci$ and $\nci$ to describe statistical independence and dependence, respectively (and $A \ci B \mid C$, to describe the statement that $A$ is independent of $B$ conditional on $C$). Note that because the measurement of the $X$ and $T$ variables precede the measurement of $Y$ (so that it is impossible for $Y$ to have a causal influence on $X$ or on $T$) we don't need to consider any statistical dependencies that are conditional on $Y$.

By inspecting the results of the following 5 statistical tests,
\begin{equation}
\begin{tabular}{lcl}
$H^1_0: \;\; T \ci X$ & vs & $H^1_1: \;\; T \nci X$, \\
$H^2_0: \;\; Y \ci X$ & vs & $H^2_1: \;\; Y \nci X$, \\
$H^3_0: \;\; Y \ci T$ & vs & $H^3_1: \;\; Y \nci T$, \\
$H^4_0: \;\; Y \ci X \mid T$ & vs & $H^4_1: \;\; Y \nci X \mid T$, \\
$H^5_0: \;\; Y \ci T \mid X$ & vs & $H^5_1: \;\; Y \nci T \mid X$, \\
\end{tabular}
\label{eq:citests}
\end{equation}
we are able to determine which among models $\{M_1, M_4\}$, $\{M_2, M_5\}$, $\{M_3, M_6\}$, $M_7$, $M_8$, and $M_9$, are supported by the data. For instance, the rejection of $H^1_0$, $H^2_0$, $H^3_0$, and $H^4_0$ together with the acceptance of $H^5_0$, indicates that the data supports $M_1$ or $M_4$, and that the association observed between $X$ and $Y$ might be due to a treatment effect. Similarly, the rejection of the $H^3_0$ and $H^5_0$ together with the acceptance of $H^1_0$, $H^2_0$, and $H^4_0$ indicates that the data supports $M_8$. It is very important, nonetheless, to keep in mind that the results of the conditional independence tests are only consistent with the causal models in (\ref{eq:models}) under the assumption that there are no unmeasured confounders. Hence, the proposed approach can only detect putative treatment and time-of-the-day effects. It is possible that, in reality, there are no treatment or time-of-the-day effects and the associations between the $\{X,T,Y\}$ measurements are actually due to unmeasured confounders. For instance, the results of the conditional independence tests consistent with $\{M_1, M_4\}$ are also consistent with $M_a$ in (\ref{eq:modelsH}), where $H$ represents an unmeasured confounder (other than short term cyclic confounders, such as circadian rhythms and daily routine schedules, for which the recorded time-of-the-day works as a surrogate variable). Similarly, test results consistent with models $\{M_2, M_5\}$, $\{M_3, M_6\}$, $M_7$, $M_8$, and $M_9$, are also consistent with $M_b$, $M_c$, $M_d$, $M_e$, and $M_f$, respectively.

\begin{equation}
{\scriptsize
\xymatrix@-0.5pc{
 & *+[F-:<10pt>]{T} & && & *+[F-:<10pt>]{T} \ar[dl] & H \ar[l] \ar[d] && H \ar[r] \ar[d] & *+[F-:<10pt>]{T} & H \ar[l] \ar[d] \\
 *+[F-:<10pt>]{X} \ar[ur] && *+[F-:<10pt>]{Y} && *+[F-:<10pt>]{X} && *+[F-:<10pt>]{Y} && *+[F-:<10pt>]{X} && *+[F-:<10pt>]{Y} \\
_{M_a} & H \ar[lu] \ar[ru] &&& _{M_b} &&&& _{M_c} & H \ar[lu] \ar[ru] & \\
 & *+[F-:<10pt>]{T} & && & *+[F-:<10pt>]{T} & H \ar[l] \ar[d] & & & *+[F-:<10pt>]{T} & H \ar[l] \ar[d] \\
 *+[F-:<10pt>]{X} && *+[F-:<10pt>]{Y} && *+[F-:<10pt>]{X} && *+[F-:<10pt>]{Y} && *+[F-:<10pt>]{X} && *+[F-:<10pt>]{Y} \\
_{M_d} & H \ar[lu] \ar[ru] &&& _{M_e} &&&& _{M_f} & H \ar[lu] \ar[ru] & \\
}}
\label{eq:modelsH}
\end{equation}

In addition to the assumption of no unmeasured confounders, the above approach also assumes the standard Markov properties for directed acyclic graphs\cite{Lauritzen1996}, and the faithfulness of the probability distribution to the graph structure\cite{Spirtes2000}.

\subsection{Accounting for serial association in the data}

In practice, we are interested in evaluating medication response at the personalized level and we analyze the data from each participant separately. In this setting, the data from each participant corresponds to a time series of treatment, time-of-the-day, and outcome variables and it is natural to expect a serial correlation structure in the data. Therefore, the causal graphs that we are actually comparing are slightly more complicated than the graphs shown in (\ref{eq:models}). For instance, the model,
\begin{equation}
{\scriptsize
\xymatrix@-0.5pc{
\ldots & *+[F-:<10pt>]{X_{t-1}} \ar[dd] \ar[dr] & & *+[F-:<10pt>]{X_{t}} \ar[dd] \ar[dr] & & *+[F-:<10pt>]{X_{t+1}} \ar[dd] \ar[dr] & & \ldots \\
\ldots & & *+[F-:<10pt>]{T_{t-1}} \ar[dl] & & *+[F-:<10pt>]{T_{t}} \ar[dl]  & & *+[F-:<10pt>]{T_{t+1}} \ar[dl] & \ldots \\
\ldots & *+[F-:<10pt>]{Y_{t-1}} & & *+[F-:<10pt>]{Y_{t}} & & *+[F-:<10pt>]{Y_{t+1}} & & \ldots \\
\ldots \ar[r] & \epsilon_{t-1} \ar[u] \ar[rr] & & \epsilon_{t} \ar[u] \ar[rr] & & \epsilon_{t+1} \ar[u] \ar[rr] & & \ldots \\
}}
\label{eq:tsmodels}
\end{equation}
represents a dynamic version of the causal graph $M_3$ in (\ref{eq:models}), where we assume that the serial correlation structure (that arises from the fact the the data comes from the same participant) is represented by the autoregressive structure of the residual error terms, $\epsilon_t$. (Note that the figure depicts a simple autoregressive serial association of order 1 only for illustrative purposes. In practice, the residual correlation structure is unknown and can be much more complicated.) The serially associated residual terms capture all factors that are not accounted by the treatment or time-of-the-day but that still influence the outcome variable over time. For instance, the performance of the participant (measured by the outcome variable) depends on the participant's current physiological state. Hence, it is reasonable to expect that the participant's performance today was not very different from its performance yesterday, and will not be very different from its performance tomorrow. However, over a longer period of time, and as the disease progresses, it is likely that the participant's performance will change due to a more degraded physiological state. Most importantly, observe that the same 5 conditional independence tests can still be used to distinguish between the dynamic versions of models $M_1$ to $M_9$ (under the additional assumption that the time series is stationary).

In our analyses, we adopt 3 distinct regression based approaches (which account for the serial correlation structures for the residuals in different ways). The first is a simple linear regression approach where we use standard t-tests for carrying out the 5 conditional independence tests in (\ref{eq:citests}) based on 4 linear regression model fits,
\begin{align}
T \, &= \, \mu \, + \, \beta_{T,X} \, X \, + \, \epsilon~, \label{eq:lm1}\\ \nonumber
Y \, &= \, \mu \, + \, \beta_{Y,X} \,  X \, + \, \epsilon~, \\ \nonumber
Y \, &= \, \mu \, + \, \beta_{Y,T} \,  T \, + \, \epsilon~, \\ \nonumber
Y \, &= \, \mu \, + \, \beta_{Y,X \mid T} \,  X \, + \, \beta_{Y,T \mid X} \,  T \, + \, \epsilon~, \nonumber
\end{align}
where the conditional independence tests in (\ref{eq:citests}) are performed by testing,
\begin{equation}
\begin{tabular}{lcl}
$H^1_0: \;\; \beta_{T,X} = 0$ & vs & $H^1_1: \;\; \beta_{T,X} \not= 0$, \\
$H^2_0: \;\; \beta_{Y,X} = 0$ & vs & $H^2_1: \;\; \beta_{Y,X} \not= 0$, \\
$H^3_0: \;\; \beta_{Y,T} = 0$ & vs & $H^3_1: \;\; \beta_{Y,T} \not= 0$, \\
$H^4_0: \;\; \beta_{Y,X \mid T} = 0$ & vs & $H^4_1: \;\; \beta_{Y,X \mid T} \not= 0$, \\
$H^5_0: \;\; \beta_{Y,T \mid X} = 0$ & vs & $H^5_1: \;\; \beta_{Y,T \mid X} \not= 0$. \\
\end{tabular}
\label{eq:citestsbetas}
\end{equation}
We employ the \texttt{lm} function of the R software\cite{Rcoreteam2014} base distribution for these analyses. Note that this approach naively assumes that the residuals of the linear regression fits are uncorrelated. Whether the serial association structure of the residuals impact the t-tests depends on whether the study participant performs the unmedicated and medicated activity tasks in a paired or un-paired (and random) fashion over time. We describe this point in more detail at the end of this section.

The second approach is based on regression with ARIMA errors modeling, where we basically fit the same 4 regression models in equations (\ref{eq:lm1}), but where the serial association of the residual errors are modeled according to an ARIMA (autoregressive integrated moving average) \cite{boxjenkins1994} process. Because the residual correlation structure is unknown, we employ the \texttt{auto.arima} function of the \texttt{forecast} R package\cite{forecast2008} in order to first select the autoregressive, moving average, and differencing orders of the models that are used to test the hypothesis in (\ref{eq:citestsbetas}). One caveat of this approach is that the ARIMA modeling assumes the data is equally spaced, what is not true in our application.

The third approach is based on regression modeling with heteroscedasticity, and autocorrelation consistent (HAC) covariance matrix estimation. Non-parametric and kernel based HAC estimators are able to account for heteroscedasticity, and autocorrelation of unknown form, and can be used to construct statistical tests that are robust to violations of homoscedasticity and independent error assumptions. Here, we adopted the Newey-West HAC estimator\cite{NeweyWest1987}, using Bartlett kernel, and the automatic bandwidth selection procedure described in reference\cite{NeweyWest1994}, and implemented in the \texttt{sandwich} R package\cite{Zeileis2004}, in order to construct robust t-tests for the same 4 regression models in equations (\ref{eq:lm1}). While HAC estimation still assumes the data is equally spaced, it has been shown that application of the Newey-West estimator to time series with missing data (and, hence, unequally spaced) still generates asymptotically consistent estimates of the covariance matrix, as well as, reasonable performance in finite sample simulation studies\cite{DattaDu2012,RhoVogelsang2014}.

\subsection{Union-intersection tests for putative treatment effects and putative ``time-of-the-day" effects}

In the previous section we described how to test for putative treatment and time-of-the-day effects for a single feature (outcome variable). In practice, however, we have multiple features and need to combine them into a single decision procedure. Here, we describe union-intersection (UI) tests for combining the feature specific tests into a single testing procedure.

Explicitly, suppose we have $p$ features indexed from $k = 1, \ldots, p$. The UI-test for a putative treatment effect is constructed by combining the feature specific tests,
\begin{equation}
H_{0k}: \beta_{Y,X \mid T} = 0  \hspace{0.5cm} \mbox{vs} \hspace{0.5cm} H_{1k}: \beta_{Y,X \mid T} \not= 0~,
\label{eq:condnull}
\end{equation}
or,
\begin{equation}
H_{0k}: \beta_{Y,X} = 0  \hspace{0.5cm} \mbox{vs} \hspace{0.5cm} H_{1k}: \beta_{Y,X} \not= 0~,
\label{eq:uncondnull}
\end{equation}
into a single test,
\begin{equation}
H_0: \; \cap_{k = 1}^{p} H_{0k} \hspace{0.5cm} \mbox{vs} \hspace{0.5cm} H_1: \; \cup_{k = 1}^{p} H_{1k}~,
\label{eq:uitest}
\end{equation}
where we use the time-of-the-day adjusted test in (\ref{eq:condnull}) when the data associated with feature $k$, $\{X, T, Y_k\}$, is consistent with the models $M_2$, $M_3$, $M_5$, $M_6$, $M_8$, and $M_9$ in (\ref{eq:models}), and the un-adjusted test in (\ref{eq:uncondnull}) otherwise. Note that because we are interested in detecting the (putative) direct effect of the treatment on the outcome, the choice to adjust or not for the time-of-the-day variable is tailored to the DAG structure (since the direct effect of $X$ on $Y$, implied by a DAG, corresponds to the effect of $X$ on $Y$ conditional on all parents of $Y$, other than $X$). (Observe, as well, that, at times, the conditional independence patterns in the data might not be consistent with any of the 9 models in (\ref{eq:models}), due to random noise or confounder induced biases, such as, M bias\cite{Greenland2003}. In these situations we don't adjust for the time-of-the-day as well.)

Similarly, the UI-test for a putative time-of-the-day effect is built by combining the feature specific tests,
\begin{equation}
H_{0k}: \beta_{Y,T \mid X} = 0  \hspace{0.5cm} \mbox{vs} \hspace{0.5cm} H_{1k}: \beta_{Y,T \mid X} \not= 0~,
\label{eq:condnulltod}
\end{equation}
or,
\begin{equation}
H_{0k}: \beta_{Y,T} = 0  \hspace{0.5cm} \mbox{vs} \hspace{0.5cm} H_{1k}: \beta_{Y,T} \not= 0~,
\label{eq:uncondnulltod}
\end{equation}
into the single test where we use the treatment adjusted test in (\ref{eq:condnulltod}) when the data is consistent with models $M_1$, $M_3$, $M_4$, $M_6$, $M_7$, and $M_9$, and the un-adjusted test in (\ref{eq:uncondnulltod}) otherwise.

Described in words, the UI-test for putative treatment (time-of-the-day) effect compares the null hypothesis of no putative treatment (time-of-the-day) effect for all features, against the alternative that there is a putative treatment (time-of-the-day) effect for at least one of the features. Under this test, we reject the null if the p-value of at least one of the feature-specific tests is small. Hence, the p-value for the UI-test corresponds to the smallest p-value (across all $p$ features) after multiple testing correction. Note, nonetheless, that because the covariate adjustment in the UI-tests are tailored to inferred DAG structures, the multiple testing correction needs to account for this as well. Hence, the correction is based on $4 \, p + 1$ tests, since for each one of the $p$ features the determination of whether the data supports any of the models in (\ref{eq:models}) is based on the 5 tests in equation (\ref{eq:citestsbetas}), but where the first one ($H^1_0: \beta_{T,X} = 0$ vs $H^1_1: \beta_{T,X} \not= 0$) is the same across all $p$ features.

Finally, note that the UI-tests can be constructed using the output of any of the 3 linear regression approaches (standard, ARIMA errors, and Newey-West) described in the previous section.

\subsection{Data processing}


For each participant, the data from each extracted feature was separately de-trended with a lowess smoother (so that our feature data, actually corresponds to the residuals of a lowess fit to the data point collection index). The data was also transformed to an approximately normal distribution using a rank-quantile transformation,
\begin{equation}
\Phi^{-1} \left( \frac{r_i - 0.5}{n} \right)~,
\end{equation}
where $\Phi()$ represents the cumulative density function of the standard normal random variable, $r_i$ represents the rank of the outcome value, $y_i$, and $n$ represents the number of outcome data points.

\subsection{On the validity of t-tests in the presence of serial correlation}

Whether residual autocorrelation impacts the type I error rates of t-tests depends on whether a participant performs the activity tasks in a paired or unpaired (and close to random) fashion. For instance, if a participant tends to perform both the unmedicated and medicated activity tasks every day (so that the data is paired by day), the residual autocorrelation can have a strong impact on the t-test p-values. In the context of paired time series, it has been shown that for positive serial correlation the F-test distribution (and, hence, the equivalent t-test in our binary treatment case) has a thicker upper tail than when the serial correlation is zero, while for negative serial correlation the upper tail is thinner\cite{McGregorBabb1989}. As a consequence, the t-tests are conservative in the presence of positive autocorrelation (i.e., the p-values tend to be larger than they should), and anti-conservative in the presence of negative autocorrelation (i.e., the p-values tend to be smaller than they should). On the other hand, if a participant tends to perform the tasks in an unpaired fashion (i.e., the before and an after medication tasks are not performed on the same day), with no particular structure about the order of the before/after tasks, then the presence of residual autocorrelation does not impact the p-value of a t-test, since the group labels (before/after medication) are exchangeable under the null hypothesis of no putative medication effect. (For further details see reference\cite{chaibubneto2016}, where it was implicitly assumed that most participants performed the before/after medication tasks in an unpaired and mostly random fashion).

\subsection{Results}

One of the goals of the mPower study is to assess if data collected by smartphones can be used to evaluate the efficacy of a new drug or intervention. To this end, we investigated whether participants showed a response to dopaminergic medication by comparing their performance in activity tasks performed before the participant has taken medication versus after medication. The longitudinal data of each participant was analyzed separately (as we are interested in determining if a given participant is responding or not to medication, as opposed to determining if dopaminergic drugs have an effect on the study cohort). For each activity task, we restricted the analyses to participants that performed at least 15 tasks before taking medication and 15 tasks after medication (106, 99, 58, and 57, participants for the tapping, voice, walk, and rest tasks, respectively).

\begin{figure}[!h]
\begin{center}
\includegraphics[width=\linewidth]{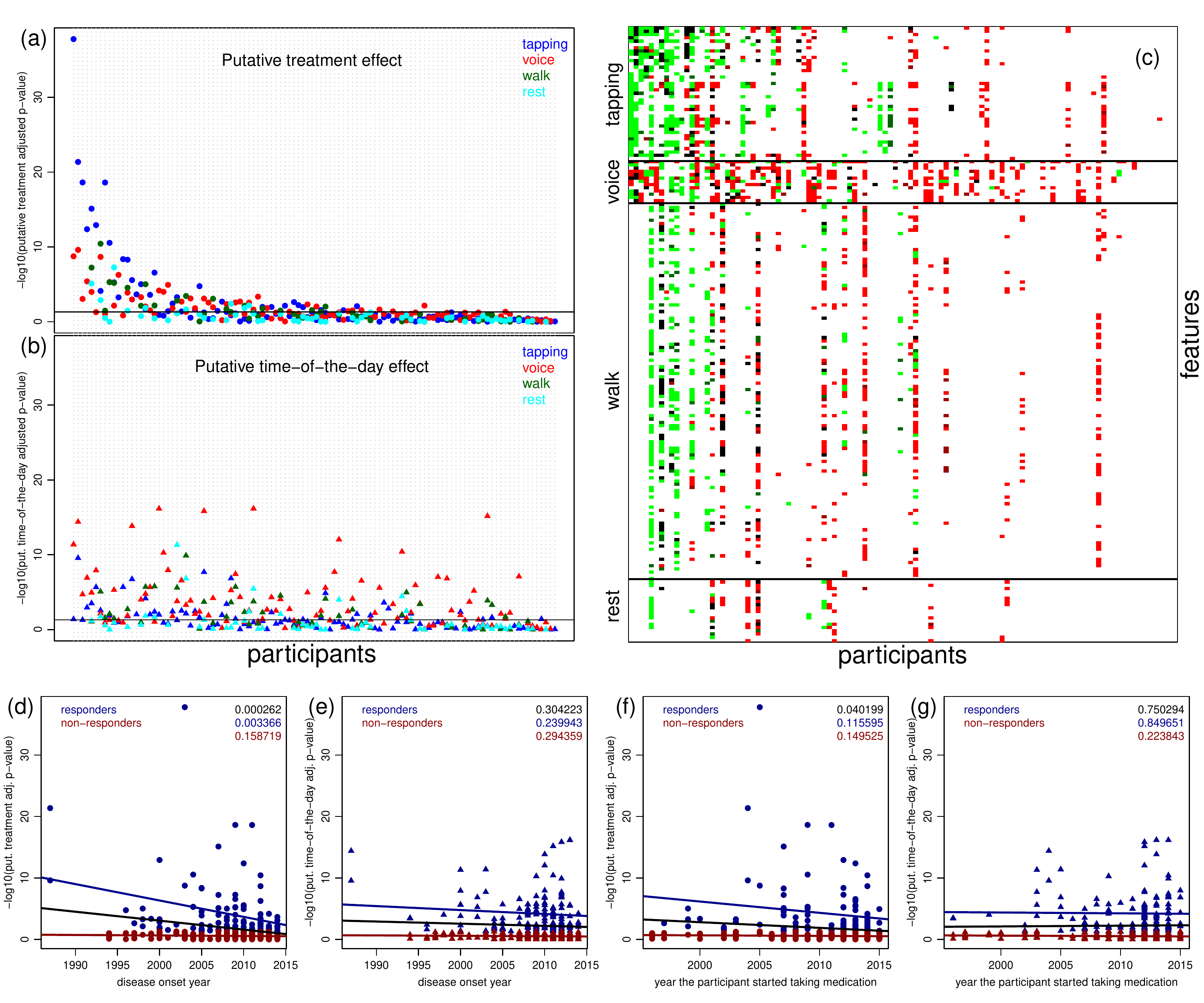}
\caption{\textbf{Personalized response to medication (standard linear regression).} Panels a and b show the adjusted p-values (in -log$_{10}$ scale) for the union-intersection tests for putative medication effects and putative time-of-the-day effects, respectively. The black line corresponds to a p-value threshold of 0.05. Panel c shows a matrix with the results of our medication vs time-of-the-day disentangling approach for all features (rows) and participants (columns). Green and red cells indicate, respectively, putative treatment and putative time-of-the-day effects, while black cells indicate both effects. White cells represent non-significant effects or missing data. The order of the participants in the x-axis of panels a, b, and c is the same, with the participants sorted according to the average putative treatment effect across the 4 activity tasks. Panels d and e show scatter plots of the disease onset year by adjusted p-values for putative medication and putative time-of-the-day effects, respectively. The blue dots represent the responders (i.e., participants that showed a significant putative medication effect in panel d, and putative time-of-the-day effect in panel e). The blue and red regression lines were fitted to the blue and red points, respectively, while the black regression line was computed using all points. The p-values for testing for zero slope of the regression lines are provided in the top right corner of the panels. Results were based on the data pooled across the 4 activity tasks. Panels f and g show analogous plots for the association with the year that the participant started taking medication.}
\label{fig:nof1Lm}
\end{center}
\end{figure}

Our analyses suggests that approximately 37\%, 42\%, 33\%, and 16\% of the participants showed a putative medication response (for at least one of the extracted features) in the tapping, voice, walk, and rest tasks, respectively, according to the union-intersection test for putative medication response based on naive linear regression models with Benjamini-Hochberg multiple testing correction at 5\% FDR (\ref{fig:nof1Lm}a).

Figure \ref{fig:nof1Lm}a shows that the participant's responses are highly individualized, with different participants showing strong putative medication effects in different activity tasks, and often times showing a response in one activity but not in another. We point out, nonetheless, that this observation is not surprising since Parkinson disease shows a wide spectrum of symptoms, and different patients show different sets of symptoms.

Figure \ref{fig:nof1Lm}b suggests that approximately 40\%, 69\%, 47\%, and 19\% of the participants showed a potential time-of-the-day response in the tapping, voice, walk, and rest tasks, respectively (according to the analogous union-intersection test for putative time-of-the-day effects, described above). Overall, we observed a larger proportion of putative time-of-the-day effects in all activities, but especially so for the voice task.

Figure \ref{fig:nof1Lm}c presents the results of our disentangling approach for each participant/feature combination. Green, red, and black cells indicate, respectively, putative treatment, putative time-of-the-day, and both putative treatment and time-of-the-day effects, while white cells represents non-significant effects or missing data (some participants didn't provide any data for some of the activity tasks). It is interesting to note that except for the voice features, where putative time-of-the-day effects are very common and some participants tended to show a somewhat balanced number of green and red cells, most participants tended to be dominated by either green or red cells. In particular, the participants in the left side of the matrix tended to show a higher proportion of putative medication effects (green cells). The consistent detection of putative medication response across several features for the same participant provides additional evidence that the effect might be real. (The less consistent results and larger proportion of time-of-the-day effects in the voice features, might be due to background noise.)

Figure \ref{fig:nof1Lm}d shows a statistically significant association (p$<$0.0005) between the p-values (in -log$_{10}$ scale) of the union-intersection test for putative treatment effects and the disease onset year of the study participants, showing that participants living with Parkinson's disease for several years tend to show the strongest differences in performance before and after taking medication. This finding agrees with what is observed in clinic, where it gets more difficult to control Parkinson's disease symptoms with dopaminergic medication as time goes by. As a sanity check, we also inspected the association between putative time-of-the-day effects and disease onset year (Figure \ref{fig:nof1Lm}e). As expected, no association was found (p$<$0.304). Figures 2f and 2g show analogous results for the association with the year that the participant started taking medication. These results provide further evidence that the observed putative medication effects might be real.

All results presented above were based on naive linear regression model fits, where we ignored the time series structure of the data. However, as pointed out above, whether residual autocorrelation impacts the type I error rates of t-tests depends on whether a participant performs the activity tasks in a paired or unpaired (and close to random) fashion.

\begin{figure}[!h]
\begin{center}
\includegraphics[width=\linewidth]{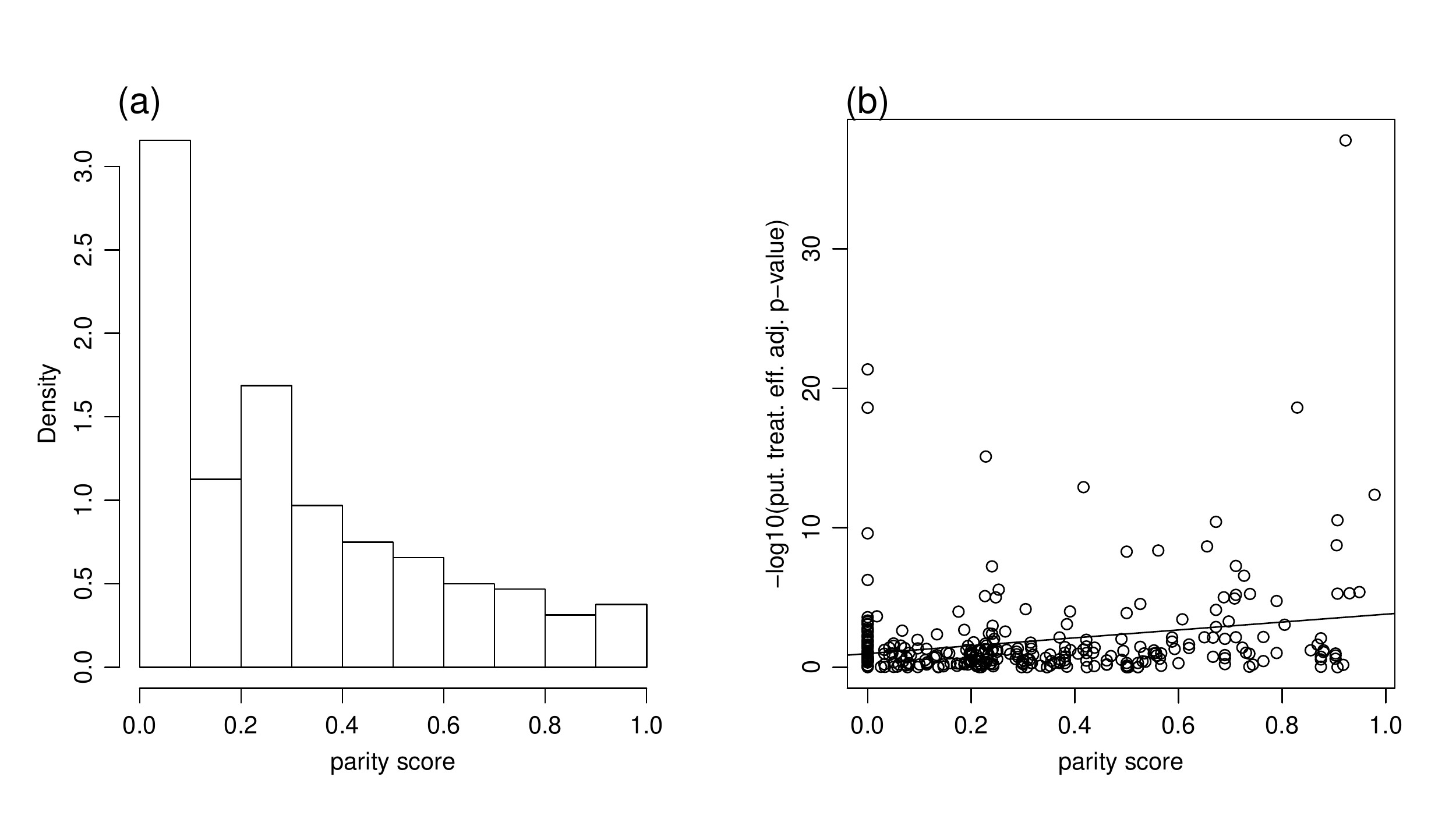}
\caption{\textbf{Before/after medication patterns.} Panel a shows the distribution of the ``parity score" of all participants, across all activity tasks. The parity score was defined as the proportion of days where the participant performed an activity tasks before and after taking medication on the same day. Panel b shows a scatter plot of the parity score versus the putative treatment effect p-values (in -log$_{10}$ scale) from the linear regression models used in the analysis presented in Figure \ref{fig:nof1Lm}.}
\label{fig:parityscore}
\end{center}
\end{figure}

In the mPower study, it is the participant who decides whether he/she will perform the activity before or after taking medication every time the participant performs a task. Inspection of the mPower's before/after label data shows a wide range of patterns ranging from completely unpaired (i.e., never performing a before and an after medication task on the same day) to almost perfectly paired (Figure \ref{fig:parityscore}a). Because some of the participants with the most significant p-values in the standard linear regression approach tended to perform the activity tasks in a more paired fashion (Figure \ref{fig:parityscore}b), it is important to investigate the residual autocorrelation of the linear regression models in order to assess if the p-values have been artificially inflated by negative autocorrelations.

\begin{figure}[!h]
\begin{center}
\includegraphics[width=\linewidth]{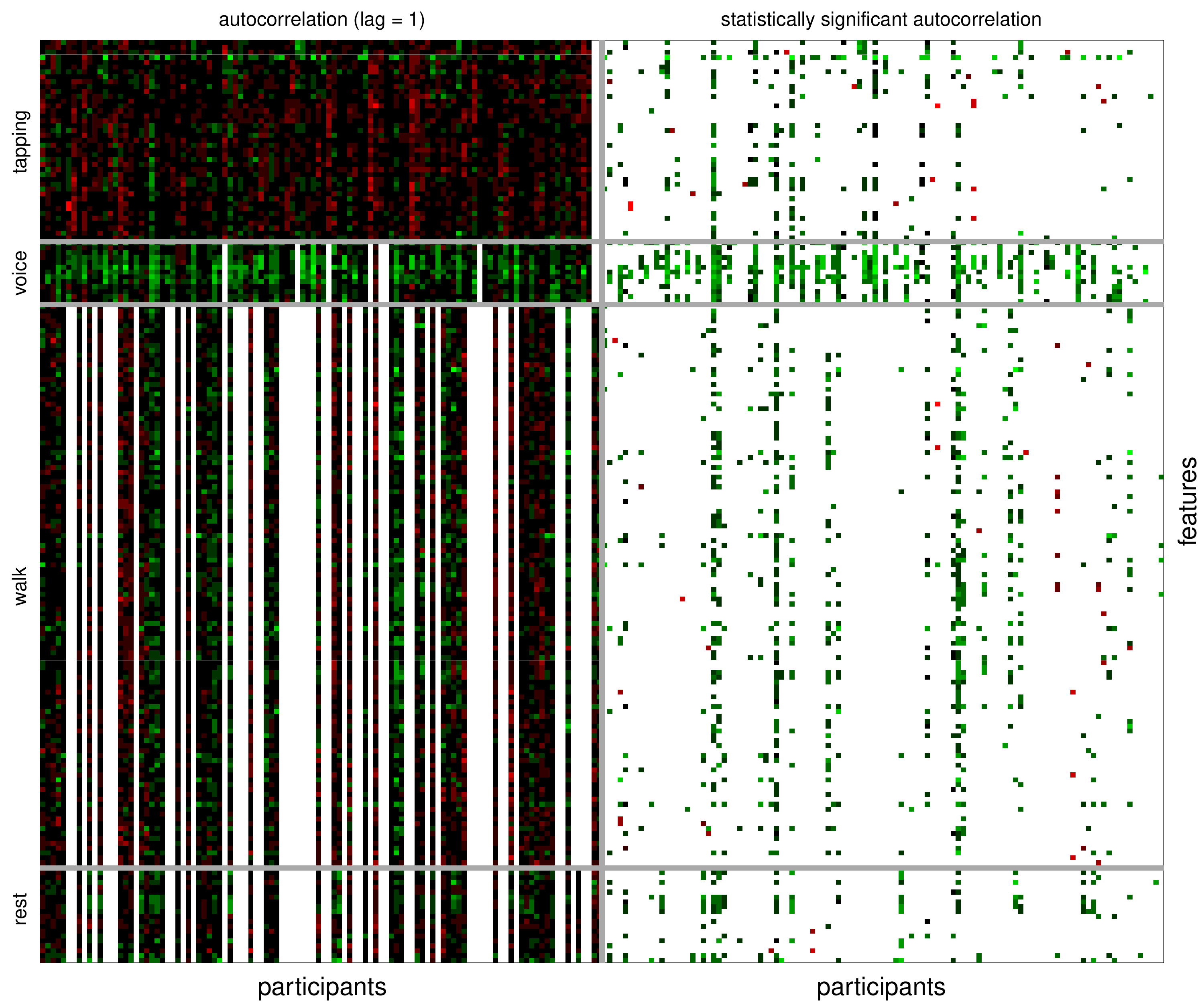}
\caption{\textbf{Residual autocorrelation of the linear regression model fits.} The heatmap in the left shows the autocorrelation (lag equal to 1) of the residuals of the linear regression model fits for each feature and participant combination. Red and green represents, respectively, negative and positive autocorrelation. White represents missing data (as some participants did not provide data for all activity tasks). The heatmap in the right shows only the autocorrelation values that were statistically different from zero according to multiple testing corrected Ljung-Box tests (based on lag equal to 1) at a significance threshold of 0.05. (The autocorrelation values that were not statistically significant are shown in white together with the missing data.) Only about 3.7\% of the statistically significant autocorrelations were negative.}
\label{fig:residautocor}
\end{center}
\end{figure}

Inspection of the residual autocorrelation structure shows, nonetheless, that most of the statistically significant autocorrelations are positive (Figure \ref{fig:residautocor}). This observation suggests that the results from the naive linear regression analyses are likely conservative, and that more sophisticated analyses accounting for autocorrelation might be better powered to detect putative medication (and putative time-of-the-day) effects.

Re-analysis of the data using regression with ARIMA errors and regression with Newey-West HAC covariance estimation confirms that modeling of the autocorrelation structure improves the power to detect putative treatment effects (Table \ref{tab:treatment}) and putative time-of-the-day effects (Table \ref{tab:tod}).

\begin{table}[!h]
\begin{center}
\begin{tabular}{lccccc}
\hline
regr. approach  & tapping & voice & walk & rest & any activity \\
\hline
standard (lm) & 37\% & 42\% & 33\% & 16\% & 60\% \\
ARIMA errors  & 42\% & 46\% & 46\% & 21\% & 69\% \\
Newey-West    & 45\% & 37\% & 42\% & 24\% & 67\% \\
\hline
\end{tabular}
\caption{\label{tab:treatment} \textbf{Proportion of participants showing statistically significant putative treatment effects, across the 3 distinct analysis approaches.} The last column shows the proportion of participants that showed a statistically significant putative medication effect in at least one of the four activity tasks.}
\end{center}
\end{table}

\begin{table}[!h]
\begin{center}
\begin{tabular}{lccccc}
\hline
regr. approach  & tapping & voice & walk & rest & any activity \\
\hline
standard (lm) & 40\% & 69\% & 47\% & 19\% & 78\% \\
ARIMA errors  & 45\% & 76\% & 53\% & 22\% & 86\% \\
Newey-West    & 57\% & 66\% & 68\% & 28\% & 83\% \\
\hline
\end{tabular}
\end{center}
\caption{\label{tab:tod} \textbf{Proportion of participants showing statistically significant putative time-of-the-day effects, across the 3 distinct analysis approaches.} The last column shows the proportion of participants that showed a statistically significant putative time-of-the-day effect in at least one of the four activity tasks.}
\end{table}

\begin{figure}[!h]
\begin{center}
\includegraphics[width=\linewidth]{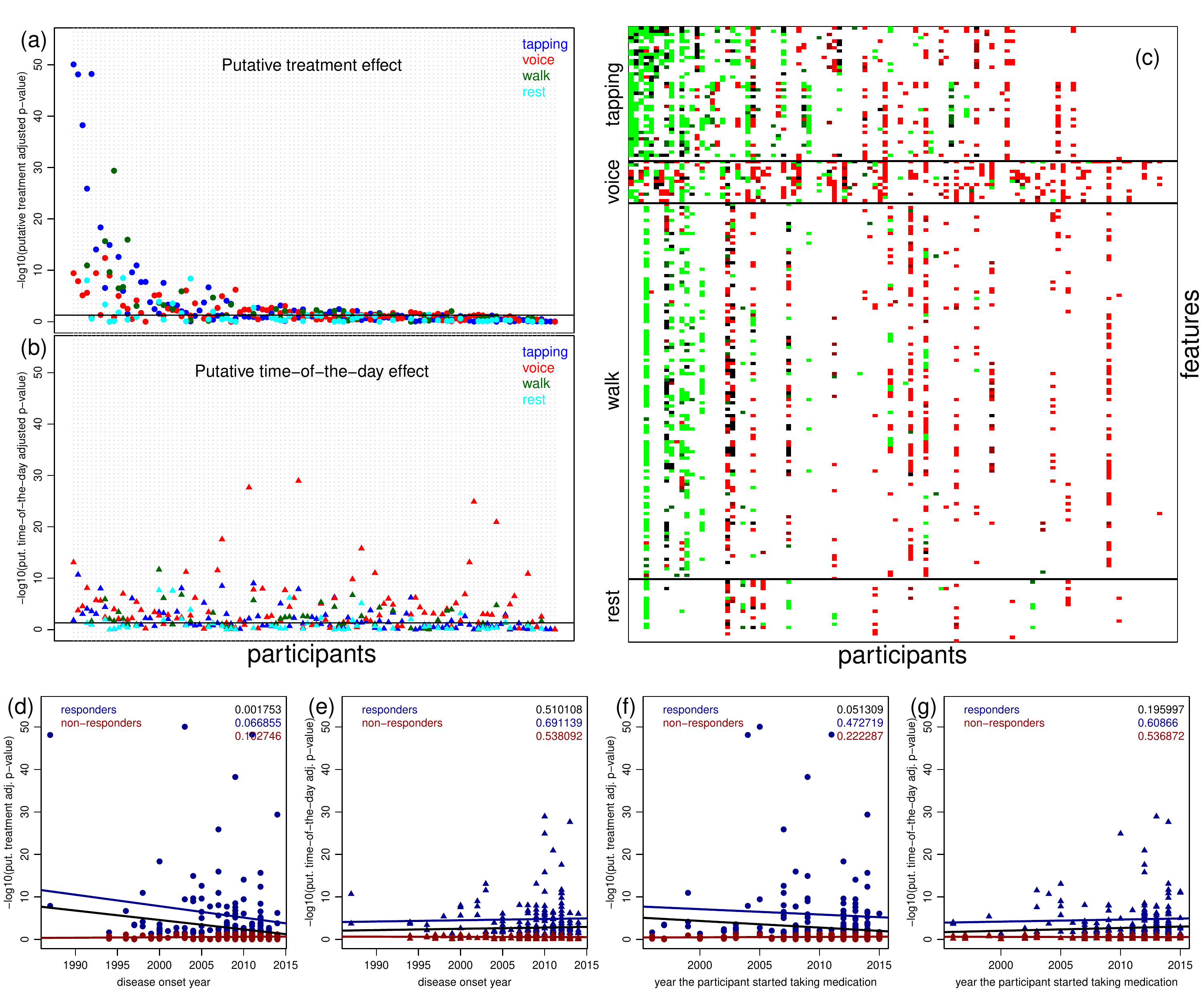}
\caption{\textbf{Personalized response to medication based on the regression with ARIMA errors approach.} Panels a and b show the adjusted p-values for the union-intersection tests for putative medication effects and putative time-of-the-day effects, respectively. The black line corresponds to a p-value threshold of 0.05. Panel c shows a matrix with the results of our medication vs time-of-the-day disentangling approach for all features (rows) and participants (columns). Green and red cells indicate, respectively, putative treatment and putative time-of-the-day effects, while black cells indicate both effects. White cells represent non-significant effects or missing data. The order of the participants in the x-axis of panels a, b, and c is the same, with the participants sorted according to the average putative treatment effect across the 4 activity tasks. Panels d and e show scatter plots of the disease onset year by adjusted p-values for putative medication and putative time-of-the-day effects, respectively. The blue dots represent the responders (i.e., participants that showed a significant putative medication effect in panel d, and putative time-of-the-day effect in panel e). The blue and red regression lines were fitted to the blue and red points, respectively, while the black regression line was computed using all points. The p-values for testing for zero slope of the regression lines are provided in the top right corner of the panels. Results were based on the data pooled across the 4 activity tasks. Panels f and g show analogous plots for the association with the year that the participant started taking medication.}
\label{fig:nof1Arima}
\end{center}
\end{figure}

\begin{figure}[!h]
\begin{center}
\includegraphics[width=\linewidth]{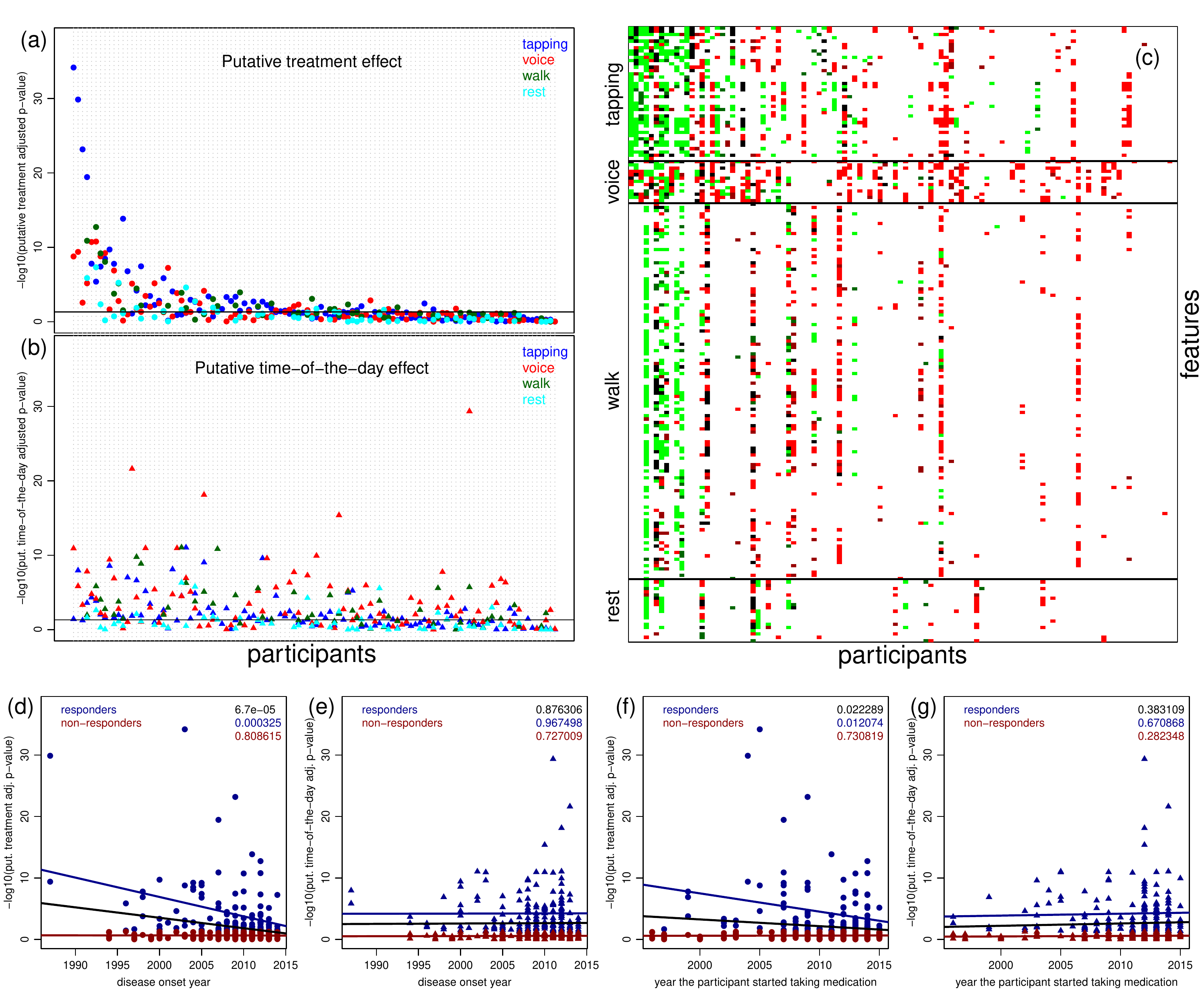}
\caption{\textbf{Personalized response to medication based on the regression with Newey-West HAC covariance estimation approach.} Panels a and b show the adjusted p-values for the union-intersection tests for putative medication effects and putative time-of-the-day effects, respectively. Panel c shows a matrix with the results of our medication vs time-of-the-day disentangling approach for all features (rows) and participants (columns). Green and red cells indicate, respectively, putative treatment and putative time-of-the-day effects, while black cells indicate both effects. White cells represent non-significant effects or missing data. The order of the participants in the x-axis of panels a, b, and c is the same, with the participants sorted according to the average putative treatment effect across the 4 activity tasks. Panels d and e show scatter plots of the disease onset year by adjusted p-values for putative medication and putative time-of-the-day effects, respectively. The blue dots represent the responders (i.e., participants that showed a significant putative medication effect in panel d, and putative time-of-the-day effect in panel e). The blue and red regression lines were fitted to the blue and red points, respectively, while the black regression line was computed using all points. The p-values for testing for zero slope of the regression lines are provided in the top right corner of the panels. Results were based on the data pooled across the 4 activity tasks. Panels f and g show analogous plots for the association with the year that the participant started taking medication.}
\label{fig:nof1Nw}
\end{center}
\end{figure}

Figures \ref{fig:nof1Arima} and \ref{fig:nof1Nw} present the same comparisons reported on Figure \ref{fig:nof1Lm}, but based on the regression with ARIMA errors and on the regression with Newey-West HAC covariance estimation approaches, respectively. Apart from the higher power to detect statistically significant putative effects, the figures illustrate that the results from all 3 analyses were largely consistent.

\section{Discussion}

This work provides two main contributions to the analysis of longitudinal data collected by smartphones in mobile health applications. In our first contribution, we proposed a novel statistical approach to disentangle putative treatment effects from putative ``time-of-the-day" effects in observational studies.

The key insight that makes the approach practical is the realization that in our mobile health application, the measurement of the treatment and time-of-the-day variables precedes the measurement of the outcome variable, so that any causal models where the outcome plays the role of a cause of the treatment or time-of-the-day variables are automatically disregarded, and it is possible to use a few conditional independence relationships to distinguish between treatment and time-of-the-day (or both) effects, irrespective of the causal relation between treatment and time-of-the-day variables. The ability to tell apart treatment and time-of-the-day effects is very important in practice, since any causal inferences about personalized treatment effects are especially vulnerable to daily cyclic confounding factors, such as circadian rhythms and daily routine activities. Application of the method to the mPower data suggests that some participants are responding to the medication, while others seem to be responding mostly to time-of-the-day. While the proposed method was motivated by the mPower study, the approach is quite general and we anticipate that the mobile health community might find this tool useful for other applications assessing personalized treatments.

Our second contribution was to show that smartphone data collected from of a given study participant can represent a ``digital signature", or a ``digital fingerprint", of the participant. As a consequence, one needs to be careful when designing the train/test split of the data used to train a case/control classifier. Explicitly, if data from each participant is included in both the training and test set, then the classifier performance can be artificially improved because it is easy to discriminate one participant from another. In order to avoid this artifact, one should instead adopt a train/test split where the data contributed by each participant is either included in the training or in the test set. Finally, we point out that while the present work has characterized the digital fingerprint issue in a Parkinson's disease application, this phenomenon might potentially show up in other applications employing data collected from highly sensitive sensors, and might represent a common artifact in mobile health studies in general.

\end{document}